\documentclass[journal]{IEEEtran}
\IEEEoverridecommandlockouts

\usepackage{cite}
\usepackage{amsmath,amssymb,amsfonts}
\usepackage{xcolor} 
\usepackage{booktabs}
\usepackage{multirow}
\usepackage{graphicx,subfigure}
\usepackage{breqn}  
\usepackage{hyperref}
\usepackage{bbm} 
\usepackage{nicematrix}
\usepackage{array} 
\usepackage{comment}

\begin{document}

\title{Self-supervised Learning for Label-Efficient Sleep Stage Classification: A Comprehensive Evaluation}

\author{Emadeldeen Eldele, Mohamed Ragab, Zhenghua Chen, Min Wu, Chee-Keong Kwoh, and Xiaoli Li
\thanks{Emadeldeen Eldele and Chee-Keong Kwoh are with the School of Computer Science and Engineering, Nanyang Technological University, Singapore (E-mail: \{emad0002, asckkwoh\}@ntu.edu.sg).}
\thanks{Mohamed Ragab and Zhenghua Chen are with the Institute for Infocomm Research (I$^2$R) and the Centre for Frontier AI Research (CFAR), Agency for Science, Technology and Research (A$^*$STAR), Singapore (E-mail: \{mohamedr002, chen0832\}@e.ntu.edu.sg).}
\thanks{Min Wu is with the Institute for Infocomm Research (I$^2$R), Agency for Science, Technology and Research (A$^*$STAR), Singapore (E-mail: wumin@i2r.a-star.edu.sg).}
\thanks{Xiaoli Li is with Institute for Infocomm Research (I$^2$R), Centre for Frontier Research (CFAR), Agency of Science, Technology and Research (A$^*$STAR), Singapore, and also with the School of Computer Science and Engineering at Nanyang Technological University, Singapore (E-mail: xlli@i2r.a-star.edu.sg).}
\thanks{First author is supported by A$^*$STAR SINGA Scholarship.}
\thanks{Min Wu is the corresponding author.}}

\maketitle

\begin{abstract}
The past few years have witnessed a remarkable advance in deep learning for EEG-based sleep stage classification (SSC). However, the success of these models is attributed to possessing a massive amount of \textit{labeled} data for training, limiting their applicability in real-world scenarios. In such scenarios, sleep labs can generate a massive amount of data, but labeling these data can be expensive and time-consuming. 
Recently, the self-supervised learning (SSL) paradigm has shined as one of the most successful techniques to overcome the scarcity of labeled data. In this paper, we evaluate the efficacy of SSL to boost the performance of existing SSC models in the few-labels regime. We conduct a thorough study on three SSC datasets, and we find that fine-tuning the pretrained SSC models with only 5\% of labeled data can achieve competitive performance to the supervised training with full labels. Moreover, self-supervised pretraining helps SSC models to be more robust to data imbalance and domain shift problems.

\end{abstract}


\section{Introduction}

Sleep stage classification (SSC) plays a key role in diagnosing many common diseases such as insomnia and sleep apnea \cite{nature_paper}. To assess the sleep quality or diagnose sleep disorders, overnight polysomnogram (PSG) readings are split into 30-second segments, i.e., epochs, and assigned a sleep stage. This process is performed manually by specialists, who follow a set of rules, e.g., the American Academy of Sleep Medicine (AASM) \cite{aasm} to identify the patterns and classify the PSG epochs into sleep stages. This manual process is tedious, exhaustive, and time-consuming.

To overcome this issue, numerous deep learning-based SSC models were developed to automate the data labeling process. These models are trained on a massive labeled dataset and applied to the dataset of interest. For example, Jadhav et al.~\cite{ssc_timeFreq} explored different deep learning models to exploit raw electroencephalogram (EEG) signals, as well as their time-frequency spectra. Also, Phyo et al.~\cite{ssc_transition} attempted to improve the performance of the deep learning model on the confusing transitioning epochs between stages. In addition, Phan et al.~\cite{sleep_transformer} proposed a transformer backbone that provides interpretable, and uncertainty quantified predictions. However, the success of these approaches hinges on a massive amount of \textit{labeled} data to train the deep learning models, which might not be feasible. In practice, sleep labs can collect a vast amount of overnight recordings, but the difficulties in labeling the data limit deploying these data-hungry models. Thus, unfortunately, the SSC works developed in the past few years have now a bottleneck: the size, quality, and availability of labeled data.

\begin{figure*}
    \centering
    \includegraphics[width=0.9\textwidth]{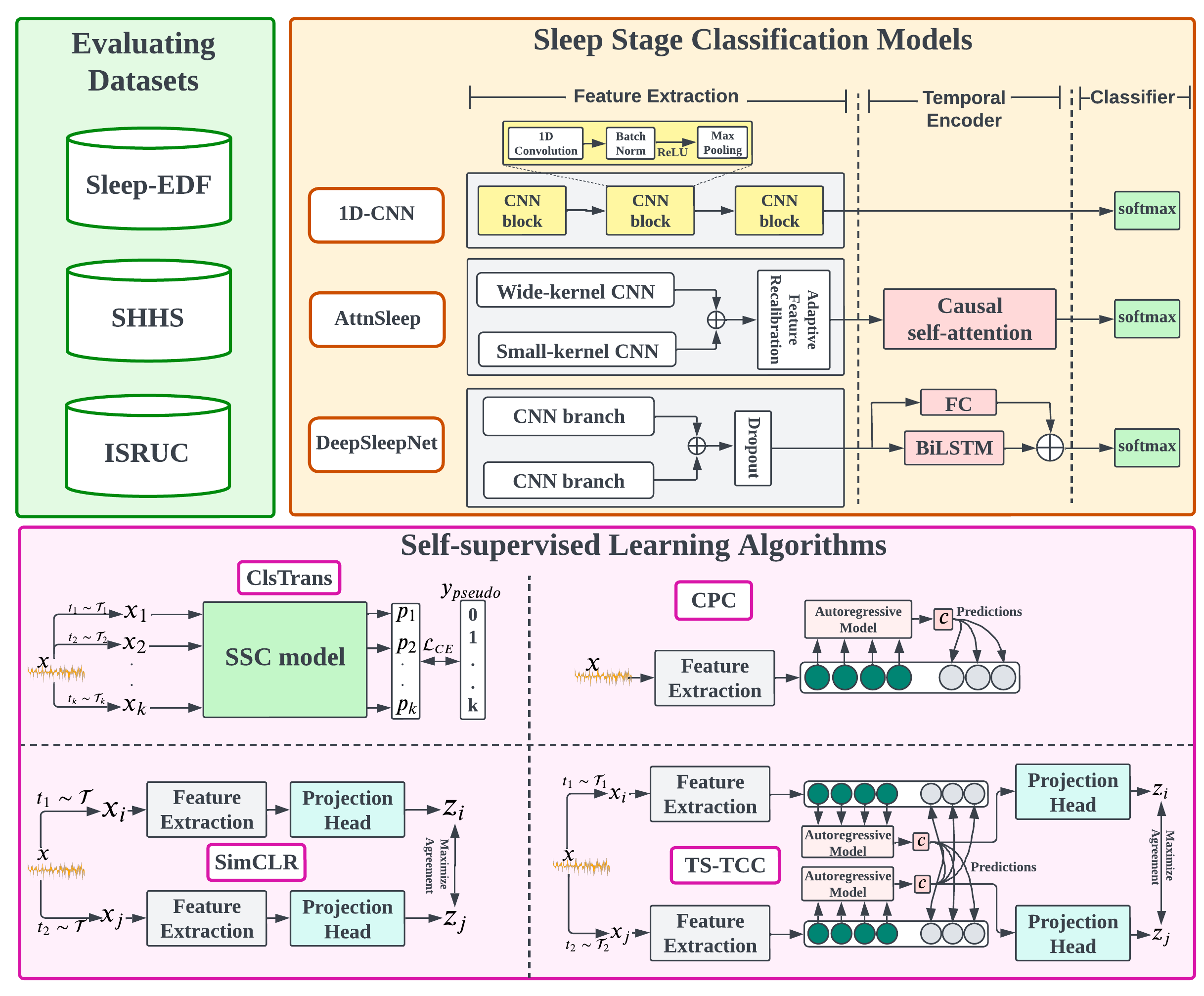}
    \caption{The architecture of our evaluation framework. We experiment with three sleep stage classification models, i.e., DeepSleepNet~\cite{deepsleepnet}, AttnSleep~\cite{attnSleep_paper}, and 1D-CNN~\cite{tstcc}. We also include four self-supervised learning algorithms, i.e., ClsTran, SimCLR~\cite{simclr}, CPC~\cite{cpc}, and TS-TCC~\cite{tstcc}. The different experiments are performed on Sleep-EDF, SHHS, and ISRUC datasets.}
    \label{fig:sleep_models}
\end{figure*}

One alternative solution to pass through this bottleneck is the self-supervised learning (SSL) paradigm, which witnessed increased interest recently due to its ability to learn useful representations from unlabeled data. In SSL, the model is pretrained on a newly defined task that does not require any labeled data, where ground-truth pseudo labels can be generated for free. Such tasks are designed to learn the model to recognize general characteristics about the data without being directed with labels.
Currently, SSL algorithms can produce state-of-the-art performance on standard computer vision benchmarks \cite{simclr,moco,simsiam,byol}. Consequently, the SSL paradigm has gained more interest to be applied for sleep stage classification problem \cite{tstcc,coSleep}.

Most prior works aim to propose novel SSL algorithms and show how they could improve the performance of sleep stage classification. Instead, in this work, our aim is to examine the efficacy of SSL paradigm to re-motivate deploying existing SSC works in real-world scenarios, where only few labeled samples are available. Therefore, we revisit a prominent subset of SSC models and perform an empirical study to evaluate their performance under the few-labeled data settings. Moreover, we explore the efficacy of different SSL algorithms on their performance and robustness. We also study the effect of sleep data characteristics, e.g., data imbalance and temporal relations, on the learned self-supervised representations. Finally, we assess the transferability of self-supervised against supervised representations and their robustness to domain shift. The overall framework is illustrated in Fig.~\ref{fig:sleep_models}. We perform an extensive set of experiments on three sleep staging datasets to systemically analyze the SSC models under the few-labeled data settings. The experimental results of this study aim to provide a solid and realistic real-world assessment of the existing sleep stage classification models.

\section{Related Work}

\subsection{Sleep Stage Classification}
A wide span of EEG-based sleep stage classification methods have been introduced in recent years. These methods proposed different architecture designs. For example, some methods adopted multiple parallel convolutional neural networks (CNNs) branches to extract better features from EEG signals \cite{deepsleepnet,attnSleep_paper,ssc_transition}. Also, some methods included residual CNN layers \cite{resnet_sleep,iinet}, while others used graph-based CNN networks \cite{graph}. On the other hand, Phan et al. \cite{seqsleepnet} proposed Long Short Term Memory (LSTM) networks to extract features from EEG spectrograms.
To handle the temporal dependencies among EEG features, these methods had different approaches. For instance, some works adopted recurrent neural networks (RNNs), e.g., bi-directional LSTM networks as in \cite{deepsleepnet,iinet,ssc_transition}. Other works adopted the multi-head self-attention as a faster and more efficient way to capture the temporal dependencies in timesteps, as in \cite{attn_residual,attnSleep_paper}.

Despite the proven performance of these architectures, they require a huge labeled training dataset to feed the deep learning models. None of these works studied the performance of their models in the few labeled data regime, which is our scope in this work.

\subsection{Self-supervised Learning Approaches}
Self-supervised learning received more attention recently because of its ability to learn useful representations from unlabeled data. The first SSL auxiliary tasks showed a big improvement in the performance of the downstream task. For example, Noroozi et al. proposed training the model to solve a jigsaw puzzle on a patched image \cite{puzzle_pretext}. In addition, Gidaris et al. proposed rotating the input images, then trained the model to predict the rotation angle \cite{rotation_pretext}.
The success of these auxiliary tasks motivated adapting contrastive learning algorithms, which showed to be more effective due to their ability to learn invariant features. The key idea behind contrastive learning is to define positive and negative pairs for each sample, then push the sample closer to the positive pairs, and pull it away from the negative pairs. In general, contrastive-based approaches rely on data augmentations to generate positive and negative pairs. For example, SimCLR considered the augmented views of the sample as positive pairs, while all the other samples within the same mini-batch are considered as negative pairs \cite{simclr}. Also, MoCo increased the number of negative pairs by keeping samples from other mini-batches in a memory bank \cite{moco}. On the other hand, some recent algorithms neglected the negative pairs and proposed using only positive pairs such as BYOL \cite{byol} and SimSiam \cite{simsiam}.

\subsection{Self-supervised learning for Sleep Staging}
The success of SSL in computer vision applications motivated their adoption for sleep stage classification.
For example, Mohsenvand et al. \cite{simclr_like_eeg} and Jiang et al. \cite{contrastive_ssc} proposed SimCLR-like methodologies and applied EEG-related augmentations for sleep stage classification. Also, Banville et al. applied three pretext tasks, i.e., relative positioning, temporal shuffling, and contrastive predictive coding (CPC) to explore the underlying structure of the unlabeled sleep EEG data \cite{uncovering_eeg}. The CPC \cite{cpc} algorithm predicts the future timesteps in the time-series signal, which motivated other works to build on it. For example, SleepDPC solved two problems, i.e., predicting future representations of epochs, and distinguishing epochs from other different epochs \cite{ssl_pred_disc}. Also, TS-TCC proposed temporal and contextual contrasting approaches to learn instance-wise representations about the sleep EEG data \cite{tstcc}. In addition, SSLAPP developed a contrastive learning approach with attention-based augmentations in the embedding space to add more positive pairs \cite{ijcai2022p537}. Last, CoSleep \cite{coSleep} and SleepECL \cite{SleepECL} are yet another two contrastive methods that exploit information, e.g., inter-epoch dependency and frequency domain views, from EEG data to obtain more positive pairs for contrastive learning.

\section{Evaluation Framework}
\subsection{Preliminaries}
In this section, we describe the SSL-related terminologies, i.e., pretext tasks, contrastive learning, and downstream tasks.

\subsubsection{Problem Formulation}
We assume that the input is single-channel EEG data in $\mathbb{R}^d$, and each sample has one label from one of $C$ classes.
The supervised downstream task has an access to the inputs and the corresponding labels, while the self-supervised learning algorithms have access only to the inputs.

The SSC networks consist of three main parts. The first is the feature extractor, which maps the input data into the embedded space $f_\phi:\mathbb{R}^d \rightarrow \mathbb{R}^{m1}$ parameterized by neural network parameters $\phi$. The second is the temporal encoder (TE), which is another intermediate network to improve the temporal representations. The TE may change the dimension of the embedded features $f_\theta:\mathbb{R}^{m1} \rightarrow \mathbb{R}^{m}$. Finally, the classifier $f_\gamma:\mathbb{R}^{m} \rightarrow \mathbb{R}^{C}$, which produces the predictions.
The SSL algorithms learn $\phi$ from unlabeled data, while fine-tuning learns $\theta$ and $\gamma$ with also updating $\phi$.

\subsubsection{Pretext tasks}

Pretext tasks refer to the pre-designed tasks to learn the model generalized representations from the unlabeled data.
Here, we describe two main types of pretext tasks, i.e., auxiliary, and contrastive tasks. 

\paragraph{Auxiliary tasks}
This category includes defining a new task along with free-to-generate pseudo labels. These tasks can be defined as classification, regression, or any others. In the context of time-series applications, a new classification auxiliary task was defined in \cite{har_ssl,ecg_emotion_rec} by generating several views to the signals using augmentations such as: adding noise, rotation, and scaling . Each view was assigned a label, and the model was pretrained to classify these transformations. This approach showed success in learning underlying representations from unlabeled data. However, it is usually designed with heuristics that might limit the generality of the learned representations \cite{tstcc}.

\paragraph{Contrastive learning} 
In contrastive learning, representations are learned by comparing the similarity between samples. In specific, we define positive and negative pairs for each sample. Next, the feature extractor is trained to achieve the contrastive objective, i.e., push the features of the sample towards the positive pairs, and pull them away from the negative pairs. These pairs are usually generated via data augmentations.
Notably, some studies \cite{simclr,emadeldeen2022catcc} relied on strong successive augmentations and found them to be a key factor in the success of their contrastive techniques. 

Formally, given a dataset with $N$ unlabeled samples, we generate two views for each sample $\mathbf{x}$, i.e., $\{\hat{\mathbf{x}}_{i}, \hat{\mathbf{x}}_{j}\}$ using data augmentations. Therefore, in a multiviewed batch with $N$ samples for each view, we have a total of $2N$ samples. Next, the feature extractor transforms them into the embedding space, and a projection head $h(\cdot)$ is used to obtain low-dimensional embeddings, i.e., $\mathbf{z}_i = h(f_\phi(\hat{\mathbf{x}}_i))$ and $\mathbf{z}_j = h(f_\phi(\hat{\mathbf{x}}_j))$. Assuming that for an anchor sample indexed $i \in I \equiv \{1 \dots 2N\}$, and $A(k) \equiv I \setminus \{k\}$. The objective of contrastive learning is to encourage the similarity between positive pairs and separate the negative pairs apart using the NT-Xent loss, defined as follows:

\begin{align}
    \mathcal{L}_{\text {NT-Xent}} &=  \frac{-1}{2N}\sum_{i \in I} \log \frac{\exp \left(\mathbf{z}_i \boldsymbol{\cdot} \mathbf{z}_j / \tau\right)}{\sum_{a \in A(i)} \exp \left(\mathbf{z}_i \boldsymbol{\cdot} \mathbf{z}_a / \tau\right)}, \label{eqn:info_nce}
\end{align}

where $\boldsymbol{\cdot}$ symbol denotes the inner dot product, and $\tau$ is a temperature parameter.

\subsubsection{Downstream tasks}
Downstream tasks are the main tasks of interest that lacked a sufficient amount of labeled data for training the deep learning models. In this paper, the downstream task is sleep stage classification, i.e., classifying the PSG epochs into one of five classes, i.e., W, N1, N2, N3, and REM. However, in general, the downstream task can be different and defined by various applications. Notably, different pretext tasks can have a different impact on the same downstream task. Therefore, it is important to design a relevant pretext task to the problem of interest, to learn better representations. Despite the numerous proposed methods in self-supervised learning, identifying the proper pretext task is still an open research question \cite{ssl_survey}.

\subsection{Sleep Stage Classification Models}
\label{sec:ssc_models}
We perform our experiments on three sleep stage classification models, i.e., DeepSleepNet \cite{deepsleepnet}, AttnSleep \cite{attnSleep_paper}, and 1D-CNN \cite{tstcc}. The architectures of these models are shown in Fig~\ref{fig:sleep_models}. Each model has its specifically-designed feature extractor, temporal encoder, and methodology to address the sleep data imbalance issue. Next, we discuss each SSC model in more details.

\subsubsection{DeepSleepNet}
DeepSleepNet consists of two parallel convolutional network branches with dropout to extract features. These features are passed to the temporal encoder that contains a Bidirectional Long Shot Term Memory (BiLSTM) network with a residual connection. To overcome the data imbalance issue in sleep data and achieve good performance in minor classes, DeepSleepNet is trained in two separate phases. In the first, the model is trained with oversampled balanced data, while in the second, the pretrained model is fine-tuned with the original imbalanced data.

\subsubsection{AttnSleep}
AttnSleep extracts features from EEG data with a multi-resolution CNN network followed by an adaptive feature recalibration module. The extracted features are then sent to a causal self-attention network to characterize the temporal relations. AttnSleep deploys a class-aware loss function to handle the class imbalance issue. This loss function assigns different weights to the data based on two factors, i.e., the distinctness of the features of each class, and the number of samples of that class in the dataset.

\subsubsection{1D-CNN}
The 1D-CNN network consists of three convolutional blocks. Each block consists of a 1D-Convolutional layer followed by a BatchNorm layer, a non-linearity ReLU activation function, and a MaxPooling layer. This architecture does not include any special component to find the temporal relations nor handle the data imbalance issue in sleep EEG data. 

In our experiments, we pretrain only the feature extractor of the three SSC models. After that, we fine-tune the whole model with the few labeled data in an end-to-end manner.

\subsection{Self-supervised Learning Algorithms}
\label{sec:ssl_techniques}
In this section, we describe the adopted SSL algorithms (see Fig.~\ref{fig:sleep_models}) in more details. We selected four algorithms that can be applied to any feature extractor design.

\subsubsection{ClsTran}
\textit{Classifying Transformations} is an auxiliary classification task, in which we first apply some transformations to the input signal. Then, we associate an automatically-generated pseudo label with each transformation. Last, we train the model to classify the transformed signals based on these pseudo labels.

Formally, let's assume a tuple of input signal and its corresponding pseudo label ($x_i, \hat{y}_i$), where $x_i$ is $i^{th}$ transformed signal, $\hat{y}_i$ is the generated pseudo label that corresponds to the $i^{th}$
transformation, and $i \in [0, T)$, $T$ is the total number of transformations. Next, the transformed signal passes through the feature extractor, the temporal encoder, and the classifier networks to generate the output probability $p_t$. Last, the model is trained to minimize a standard cross-entropy loss based on these pseudo labels: $\mathcal{L}_{\mathrm{CE}}= \sum_{t=0}^{T-1}  \mathbbm{1}_{[\hat{y} = t]} \log p_t$, where $\mathbbm{1}$ is the indicator function, which is set to be 1 when the condition is met, and set to 0 otherwise. In this work, we adopt four augmentations, i.e., negation, permutation, adding noise, and time shifting, which were adopted by previous works \cite{simclr_like_eeg,tstcc} and showed good downstream performance. More details about data augmentations are provided in Section~\ref{sec:data_augmentation} in the supplementary materials.

\subsubsection{SimCLR}
\textit{Simple framework for Contrastive Learning of Visual Representation} \cite{simclr} is a contrastive SSL algorithm that relies on data augmentations to learn invariant representations. It consists of four major components. The first is data augmentations, which are utilized to generate two correlated views of the same sample. The second is the feature extractor network that transforms the augmented views into latent space. The third is the projection head, which maps the features into a low-dimensional space. The fourth is the NT-Xent loss (Eq.~\ref{eqn:info_nce}), which aims to maximize the similarity between an anchor sample with its augmented views while minimizing its similarity with the augmented views of the other samples within the mini-batch.

\subsubsection{CPC}
\textit{Contrastive Predictive Coding} \cite{cpc} is a predictive contrastive SSL approach that learns representations of time-series signals by predicting the future timesteps in the embedding space. To do so, the feature extractor first generates the latent feature embeddings for the input signals. Next, an autoregressive model receives a part of the embeddings, i.e., the past timesteps, then generates a context vector and uses it to predict the other part, i.e., the future timesteps. CPC deploys a contrastive loss such that the embedding should be close to positive future embeddings and distant from negative future embeddings. CPC showed improved downstream performance in various time-series and speech recognition-related tasks, without the need for any data augmentation.

\begin{table*}[!htb]
\centering
\caption{Results of fine-tuning pretrained models with different SSL techniques with only 1\% of labels (Per-class performance is in terms of F1-score). \textbf{Best} results are in bold, while \underline{second best} are underlined.}
\begin{NiceTabular}{c|l|  
     *{4}{c @{\hskip 6pt}}   c *{2}{c @{\hskip 4pt}}|
     *{4}{c @{\hskip 6pt}}   c *{2}{c @{\hskip 4pt}}|
     *{4}{c @{\hskip 6pt}}   c *{2}{c @{\hskip 4pt}} }
\toprule
& & \multicolumn{7}{c|}{\textbf{DeepSleepNet}} & \multicolumn{7}{c|}{\textbf{AttnSleep}} & \multicolumn{7}{c}{\textbf{1D-CNN}} \\ \cmidrule(l){2-23} 
& Algorithm & W & N1 & N2 & N3 & REM & ACC & MF1 & W & N1 & N2 & N3 & REM & ACC & MF1 & W & N1 & N2 & N3 & REM & ACC & MF1 \\ \cmidrule(l){2-23}
 
& Supervised & 67.4 & 24.1 & 78.0 & 80.6 & 51.7 & 68.2 & 60.4 & 71.5 & 17.1 & 72.9 & 79.9 & 46.5 & 65.3 & 57.6 & 68.0 & 14.7 & 73.0 & 70.8 & 53.3 & 64.3 & 55.9 \\

& ClsTran & 60.1 & 17.2 & 69.6 & 73.1 & 48.1 & 61.2 & 53.6 &  57.4 & 11.5 & 65.5 & 64.8 & 42.2 & 57.3 & 48.3 & 72.7 & 12.7 & 76.6 & 78.2 & 47.1 & 67.9 & 57.5 \\

& SimCLR & 76.7 & \textbf{25.2} & \underline{83.0} & 79.5 & \textbf{65.7} & \underline{74.8} & \underline{66.0} & 67.9 & 18.6 & \textbf{80.0} & \underline{80.6} & \textbf{58.7} & \underline{70.5} & \underline{61.2} & \underline{80.6} & \underline{19.7} & \textbf{83.7} & \underline{84.4} & \underline{66.1} & \underline{76.4} & \underline{66.9} \\

& CPC & \textbf{78.7} & 21.3 & \textbf{84.1} & \underline{82.3} & 61.3 & \underline{74.8} & 65.5 & \underline{72.7} & \underline{20.0} & 78.3 & 78.4 & 51.6 & 68.6 & 60.2 & 80.3 & \textbf{20.0} & 81.3 & 80.1 & 59.0 & 73.3 & 64.1 \\

\multirow{-7}{*}{\rotatebox[origin=c]{90}{\textbf{Sleep-EDF}}} & TS-TCC  & \underline{78.5} & \underline{24.5} & 82.9 & \textbf{83.2} & \underline{63.8} & \textbf{75.2} & \textbf{66.6} & \textbf{77.8} & \textbf{22.1} & \underline{78.8} & \textbf{83.5} & \underline{53.9} & \textbf{71.2} & \textbf{63.2} & \textbf{82.7} & 19.2 & \underline{83.2} & \textbf{84.7} & \textbf{66.4} & \textbf{76.5} & \textbf{67.2} \\ 

\bottomrule
\toprule

& Algorithm  & W & N1 & N2 & N3 & REM & ACC & MF1 & W & N1 & N2 & N3 & REM & ACC & MF1 & W & N1 & N2 & N3 & REM & ACC & MF1 \\ \cmidrule(l){2-23} 

& Supervised    & 63.7 & 1.0 & 73.7 & 76.2 & 51.2 & 66.8 & 53.2 & 55.0 & 3.8 & 68.8 & 69.4 & 46.3 & 60.3 & 48.7 & 43.6 & 0.5 & 65.8 & 59.2 & 45.7 & 56.9 & 42.9 \\

& ClsTran       & 39.5 & 0.2 & 61.5 & 63.0 & 34.7 & 52.9 & 39.8 & 59.8 & \underline{6.7} & 65.0 & 59.5 & 45.8 & 58.5 & 47.3 & 62.9 & 0.1 & 69.8 & 69.7 & 39.9 & 62.6 & 48.5 \\

& SimCLR        & \textbf{77.3} & \textbf{7.8} & \underline{77.3} & \underline{77.8} & \underline{54.0} & \textbf{71.9} & \textbf{58.8} & 71.3 & \textbf{7.2} & \underline{74.6} & \underline{76.6} & \textbf{48.6} & \underline{68.4} & \underline{55.6} & \textbf{78.9} & 3.3 & \textbf{79.4} & \textbf{81.7} & \textbf{54.6} & \textbf{74.2} & \textbf{59.6} \\

& CPC           & 72.8 & \underline{4.2} & 76.0 & 76.6 & \textbf{54.2} & 69.5 & 56.8 & \underline{62.4} & 5.5 & 55.9 & 63.4 & 36.6 & 53.6 & 44.8 & 76.9 & \textbf{9.8} & 71.6 & 70.9 & 50.4 & 67.6 & 55.9 \\ 

\multirow{-7}{*}{\rotatebox[origin=c]{90}{\textbf{SHHS}}} & TS-TCC        & \underline{74.4} & 3.7 & \textbf{78.2} & \textbf{78.4} & 52.8 & \underline{71.2} & \underline{57.5} & \textbf{73.3} & 4.6 & \textbf{74.8} & \textbf{77.5} & \underline{48.5} & \textbf{68.9} & \textbf{55.7} & \underline{77.8} & \underline{4.8} & \underline{78.7} & \underline{80.5} & \underline{51.2} & \underline{72.9} & \underline{58.6} \\

\bottomrule
\toprule

& Algorithm  & W & N1 & N2 & N3 & REM & ACC & MF1 & W & N1 & N2 & N3 & REM & ACC & MF1 & W & N1 & N2 & N3 & REM & ACC & MF1 \\ \cmidrule(l){2-23} 

& Supervised    & 63.6 & 35.0 & 44.6 & 75.5 & \textbf{46.7} & 55.5 & 53.1 & 63.7 & \textbf{38.5} & 43.3 & 73.8 & 21.3 & 52.9 & 48.1 & 51.7 & 28.7 & 27.3 & 63.9 & 36.1 & 46.3 & 41.5 \\

& ClsTran       & 49.5 & 33.6 & 36.5 & 64.4 & 32.5 & 46.9 & 43.3 & 58.1 & 33.5 & 26.3 & 60.3 & 22.1 & 41.4 & 40.0 & 36.6 & 27.4 & 18.4 & 74.6 & 31.9 & 42.7 & 37.8 \\

& SimCLR        & 70.0 & \underline{40.2} & 49.6 & \underline{77.9} & \underline{42.6} & 58.6 & 56.1 & \underline{69.2} & \underline{37.6} & \underline{53.8} & \underline{75.9} & 30.0 & \underline{58.7} & \underline{53.3} & 76.7 & \underline{42.1} & 55.4 & 76.0 & 21.0 & 59.7 & 54.3 \\

& CPC           & \underline{76.5} & 33.7 & \textbf{62.6} & \textbf{80.3} & 40.5 & \textbf{65.1} & \underline{58.7} & \textbf{77.0} & 32.2 & \textbf{57.3} & \textbf{78.3} & \underline{31.2} & \textbf{60.6} & \textbf{55.2} & \underline{77.0} & 39.6 & \textbf{58.5} & \textbf{85.4} & \underline{37.5} & \underline{63.5} & \underline{59.6} \\ 

\multirow{-7}{*}{\rotatebox[origin=c]{90}{\textbf{ISRUC}}} & TS-TCC        & \textbf{79.5} & \textbf{44.5} & \underline{58.1} & 77.4 & 42.0 & \underline{63.6} & \textbf{60.3} & 69.0 & 36.3 & 49.0 & 73.5 & \textbf{31.8} & 54.9 & 51.9 & \textbf{82.0} & \textbf{42.5} & \underline{56.8} & \underline{84.9} & \textbf{46.4} & \textbf{65.9} & \textbf{62.5} \\

\bottomrule
\end{NiceTabular}
\label{tbl:all_results_3_datasets}
\end{table*}

\subsubsection{TS-TCC}
\textit{Time-Series representation learning via Temporal and Contextual Contrasting} \cite{tstcc} is yet another contrastive SSL approach for time-series data. TS-TCC relies on strong and weak augmentations to generate two views of an anchor sample.
Next, the feature embeddings of these views are generated. Next, similar to CPC, a part of the embeddings of each view is sent to an autoregressive model to generate a context vector. Then, the context vector generated for one augmented view is used to predict the future timesteps of the other augmented view with a contrastive loss. Therefore, it pushes the embeddings of one augmented view to the positive future embeddings of the other augmented view, and vice versa. In addition, it leverages the NT-Xent loss (Eq.~\ref{eqn:info_nce}) to maximize the agreement between the context vectors of the same sample, while maximizing it within the contexts of other samples.

\section{Experimental Setup}

\subsection{Datasets}
We evaluate the SSL algorithms on three sleep stage classification datasets, namely Sleep-EDF, SHHS, and ISRUC. These datasets have different characteristics in terms of sampling rates, EEG channels, and the health conditions of subjects. We use a single EEG channel from each dataset in our experiments following previous works \cite{deepsleepnet,attnSleep_paper}.

\subsubsection{Sleep-EDF}
Sleep-EDF dataset \cite{sleep_edf} is a public dataset that contains the polysomnography (PSG) readings of 20 healthy subjects (10 males and 10 females). In our experiments, we adopted the recordings included in the Sleep Cassette (SC) study and used the EEG data from Fpz-Cz channel with a sampling rate of 100 Hz.

\subsubsection{SHHS}
Sleep Heart Health Study \cite{shhs_ref1,shhs_ref2} is a multi-center cohort study of the cardiovascular and other consequences of sleep-disordered breathing.
The dataset is created to record the PSG readings of patients aged 40 years and older in two visits. In our experiments, we randomly chose 20 subjects from the patients during the first visit and chose the EEG channel C4-A1 with a sampling rate of 125 Hz. 

\subsubsection{ISRUC}
ISRUC dataset \cite{isruc_dataset} contains PSG recordings for human adults with different health conditions. We selected the 10 healthy subjects included in subgroup III and extracted the EEG channel C4-A1 with a sampling rate of 200 Hz. 

More details about the datasets are provided in Table~\ref{tbl:data}.

\begin{table}[!htb]
\centering
\caption{Details of the three datasets used in our experiments (each sample is a 30-second epoch). S.R. refers to the sampling rate.}
\resizebox{\columnwidth}{!}{
\begin{tabular}{@{}ccccccccc@{}}
\toprule
\textbf{Datasets} & \textbf{Channel}  & \textbf{S.R.}  & \textbf{W} & \textbf{N1} & \textbf{N2} & \textbf{N3} & \textbf{REM} & \textbf{\#Total}\\ \midrule
\multirow{2}{*}{\textbf{Sleep-EDF}} & \multirow{2}{*}{Fpz-Cz} & \multirow{2}{*}{100 Hz}
& 8285 & 2804 & 17799 & 5703 & 7717 &  \multirow{2}{*}{42308}\\
 & &  & \textit{19.6\%} & \textit{6.6\%} & \textit{42.1\%} & \textit{13.5\%} & \textit{18.2\%} \\ 
 
 \midrule
 
\multirow{2}{*}{\textbf{SHHS}} & \multirow{2}{*}{C4-A1} & \multirow{2}{*}{125 Hz} & 4741 & 783 & 8204 & 2747 & 3546 & \multirow{2}{*}{20021}  \\
 &  &  &\textit{23.7\%} & \textit{3.9\%} & \textit{40.9\%} & \textit{13.7\%} & \textit{17.8\%}\\ 

\midrule

\multirow{2}{*}{\textbf{ISRUC}} & \multirow{2}{*}{C4-A1} & \multirow{2}{*}{200 Hz} & 1817 & 1248 & 2678 & 2035 & 1111 & \multirow{2}{*}{8889}  \\
 &  &  &\textit{20.4\%} & \textit{14.0\%} & \textit{30.1\%} & \textit{22.9\%} & \textit{12.5\%}\\

\bottomrule
\end{tabular}
}
\label{tbl:data}
\end{table}

\subsection{Implementation Details}
\subsubsection{Dataset preprocessing}
For all the datasets, we apply the two preprocessing steps. First, we only considered the five sleep stages according to the AASM standard. Second, we exclude the wake periods that exceed 30 minutes before and after the sleep periods following \cite{deepsleepnet,attnSleep_paper}. We also split the subjects into five folds, and \textit{all} the upcoming experiments are performed with 5-fold subject-wise cross-validation.

\subsubsection{Training scheme}
The pretraining as well as the fine-tuning were performed for 40 epochs with a batch size of 128. The neural network weights were optimized using the Adam optimizer, with a learning rate of 1e-3 and a weight decay of 1e-4. We reported the results in terms of accuracy and macro F1-score. Our codes are built using PyTorch 1.7 and they are publicly available at \href{https://github.com/emadeldeen24/eval_ssl_ssc}{https://github.com/emadeldeen24/eval\_ssl\_ssc}.


\section{Results}

\subsection{Which SSL algorithm performs best?}
In Tables~\ref{tbl:all_results_3_datasets}, we compare the \textit{supervised} performance of the three SSC models (Section~\ref{sec:ssc_models}) against the \textit{fine-tuned} models with the four SSL algorithms (Section~\ref{sec:ssl_techniques}) using 1\% of labeled data.

We notice that self-supervised pretraining with contrastive methods ensures better performance against supervised training in the few-labeled data regime. Specifically, we find that SimCLR, CPC, and TS-TCC demonstrate remarkable performance on the three datasets. This indicates that learning invariant representations by contrastive learning can achieve good generalization on sleep datasets. Counterpart, pretraining with the auxiliary task learns poorer representations, leading to a downgraded performance except for few cases. This could be regarded to the high complexity of sleep EEG data, which does not help the model identify the difference between several augmented views.

We also conducted several experiments to assess the capability of SSL algorithms in learning temporal information, which are provided in the supplementary materials (see Section~\ref{sec:supp:te}). We find that pretrained models with CPC and TS-TCC can be robust to the existence and the type of temporal encoder while fine-tuning. The reason is that these methods rely on predicting the future timesteps in the latent space, which allows them to learn about temporal features in the EEG data.


\begin{figure}[!tb]
    \centering
    \includegraphics[width=\columnwidth]{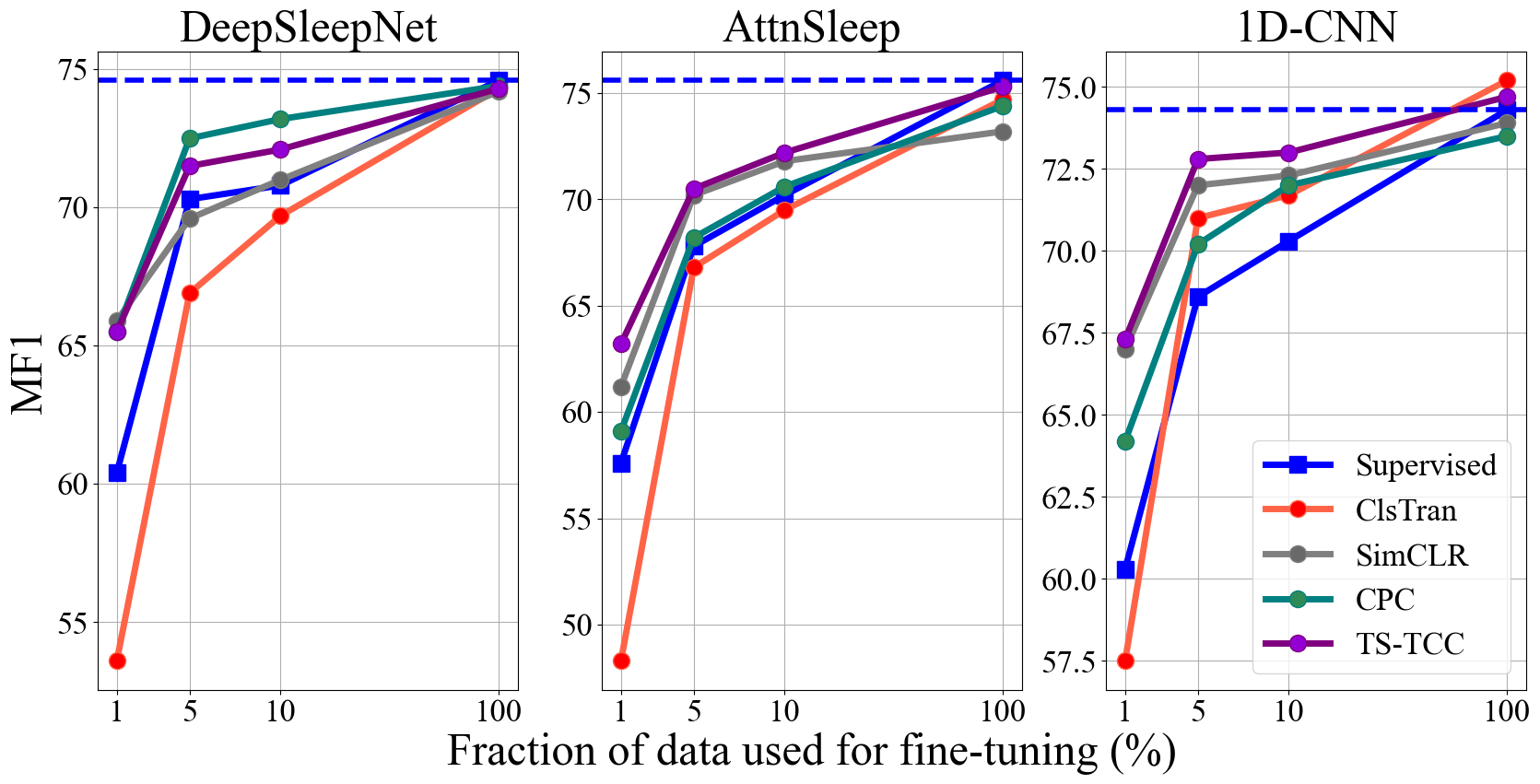}
    \caption{Fine-tuning the pretrained SSL algorithms with different fractions of labeled Sleep-EDF data.}
    \label{fig:few_lbl}
\end{figure}

\subsection{Performance Under Different Few-labels Settings} 
We study the performance of pretrained models when fine-tuned with different amounts of labeled data, i.e., 1\%, 5\%, 10\%, and 100\%. Fig.~\ref{fig:few_lbl} shows the result of these experiments on the Sleep-EDF dataset (results on SHHS and ISRUC datasets are provided in Section~\ref{sec:supp:few_lbls} in the supplementary materials).
We find that for the three SSC models, fine-tuning with 5 or 10\% of labels can achieve very close performance to the supervised training with 100\% of labels.
This demonstrates that self-supervised pretrained models yield richer embeddings than their supervised counterparts, which enhances the downstream task performance with such few labels.  

Specifically, fine-tuning CPC-pretrained DeepSleepNet with 5\% of labeled data could achieve an F1-score of 72.5\%, which is only 2.1\% less than supervised training with full labels. Also, fine-tuning TS-TCC-pretrained AttnSleep and 1D-CNN with 5\% of labeled data had a difference from the fully supervised training of 5.1\% and 1.5\% respectively.
Similarly, fine-tuning with 10\% of labeled data have even lower differences of 1.4, 3.4, and 1.3\% on DeepSleepNet, AttnSleep, and 1D-CNN respectively with fully supervised training.
These results indicate the applicability of existing SSC works in real-world scenarios provided the self-supervised pretraining. 

We also find that the gain from self-supervised pretraining tends to diminish with fine-tuning the model with the full labeled data. This observation holds for all the three SSC models on the three datasets.
Therefore, we can conclude that self-supervised pretraining can provide better regularization, reducing the overfitting problem. However, it does not improve the optimization to reduce the underfitting problem, which is aligned with the findings in \cite{9157100}.


\subsection{Comparison with Baselines}
We compare the performance of the adopted pretrained SSC models against state-of-the-art self-supervised methods proposed specifically for the sleep stage classification problem.
Table~\ref{tbl:comp_w_sota} provides the comparison results, where we show the reported results of SleepDPC~\cite{ssl_pred_disc}, CoSleep~\cite{coSleep}, and SSLAPP~\cite{ijcai2022p537} on Sleep-EDF dataset.
We compare these methods against existing SSC models with the best-performing SSL method.

Despite that experimental settings can be in favor of state-of-the-art sleep-specific SSL methods, e.g., using multiple channels and different data splits, we find that pretraining existing SSC models surpass their performance under the few labels settings. For example, we find that SleepDPC and CoSleep include two EEG channels in training, yet they achieve poor performance. Also, SSLAPP shows less performance despite that it splits the data into 80/20, which may not provide dependable conclusions as the K-fold subject-wise cross-validation. On the other hand, pretrained SSC models with a single EEG channel were able to outperform these methods in terms of both accuracy and macro F1-score. Therefore, it is important to rebirth existing SSC models with self-supervised learning to obtain comparable real-world scenario results.

\begin{table}[!tb]
\centering
\caption{Comparison between SOTA Self-supervised methods for sleep stage classification and existing fine-tuned SSC models. Experiments are applied on Sleep-EDF dataset.}
\begin{NiceTabular}{l|*{3}{c@{\hskip 3pt}}|*{2}c@{\hskip 1pt}}
\toprule
Method & Data Split & Chanels & Labels\% & ACC & MF1 \\ \midrule
SleepDPC & 20-fold CV & 2 EEG & 10 & 70.1 & 64.0 \\
CoSleep & 10-fold CV & 2 EEG & 10 & 71.6 & 55.8 \\
SSLAPP & 80/20 & EEG+EOG & 10 & 77.2 & 72.0 \\
\midrule
DeepSleepNet+CPC & 5-fold CV & 1 EEG & 10 & \textbf{81.0} & \textbf{73.2} \\
AttnSleep+TS-TCC & 5-fold CV & 1 EEG & 10 & 79.8 & 72.2 \\
1D-CNN+TS-TCC & 5-fold CV & 1 EEG & 10 & \underline{80.9} & \underline{73.0} \\ \bottomrule
\end{NiceTabular}
\label{tbl:comp_w_sota}
\end{table}



\begin{table*}[!htb]
\centering
\caption{Transferability experiment applied on Sleep-EDF dataset for 5 cross-subject scenarios. \textbf{Best} results are in bold, while \underline{second best} are underlined. Results are in terms of MF1-score.}
\begin{NiceTabular}{@{}l|
        *{4}{c @{\hskip 6pt}}   c *{1}{c @{\hskip 6pt}}|
        *{4}{c @{\hskip 6pt}}   c *{1}{c @{\hskip 6pt}}|
        *{4}{c @{\hskip 6pt}}   c *{1}{c @{\hskip 6pt}}}

\toprule
 & \multicolumn{6}{c|}{DeepSleepNet} & \multicolumn{6}{c|}{AttnSleep} & \multicolumn{6}{c}{1D-CNN} \\ \cmidrule(l){2-19} 
 & 0$\rightarrow$7 & 1$\rightarrow$8 & 2$\rightarrow$9 & 3$\rightarrow$10 & 4$\rightarrow$11 & Avg & 0$\rightarrow$7 & 1$\rightarrow$8 & 2$\rightarrow$9 & 3$\rightarrow$10 & 4$\rightarrow$11 & Avg & 0$\rightarrow$7 & 1$\rightarrow$8 & 2$\rightarrow$9 & 3$\rightarrow$10 & 4$\rightarrow$11 & Avg \\ \midrule
Supervised & 66.4 & 54.8 & 51.9 & 52.0 & 51.3 & 55.3 & \underline{65.8} & 54.0 & \underline{62.1} & \underline{57.8} & 52.2 & \underline{58.4} & 63.8 & \underline{59.8} & 54.8 & 54.1 & 46.1 & 55.7\\ 
ClsTran & \underline{69.2} & \underline{55.0} & \underline{52.7} & 45.3 & \underline{52.2} & 54.9 & 64.1 & \underline{56.7} & 56.1 & 50.3 & 50.9 & 55.6 & 64.9 & 53.6 & \underline{60.9} & 51.7 & 50.1 & 56.2 \\
SimCLR & 63.1 & 52.9 & 51.4 & \underline{52.2} & \textbf{59.7} & \underline{55.9} & 61.7 & 53.2 & \textbf{64.2} & 56.0 & 51.2 & 57.3 & 62.4 & \textbf{60.2} & 60.8 & 53.6 & \textbf{57.1} & \underline{58.8} \\
CPC & \textbf{71.8} & \textbf{61.9} & 50.5 & 49.9 & 48.5 & \textbf{56.5} & \textbf{68.6} & \textbf{62.9} & 55.8 & \textbf{60.3} & \underline{52.3} & \textbf{60.0} & \textbf{72.6} & 52.7 & 56.8 & \textbf{59.6} & 45.5 & 57.5  \\
TS-TCC &  63.0 & 51.8 & \textbf{56.9} & \textbf{53.4} & \underline{52.2} & 55.5 & 65.6 & 53.5 & 58.4 & 54.6 & \textbf{56.1} & 57.6 & \underline{71.2} & 58.2 & \textbf{61.0} & \underline{57.5} & \underline{51.4} & \textbf{59.9} \\
\bottomrule
\end{NiceTabular}
\label{tbl:transfer_learning}
\end{table*}

\subsection{Robustness of SSL against Sleep Data Imbalance}

\begin{figure}
    \centering
    \includegraphics[width=\columnwidth]{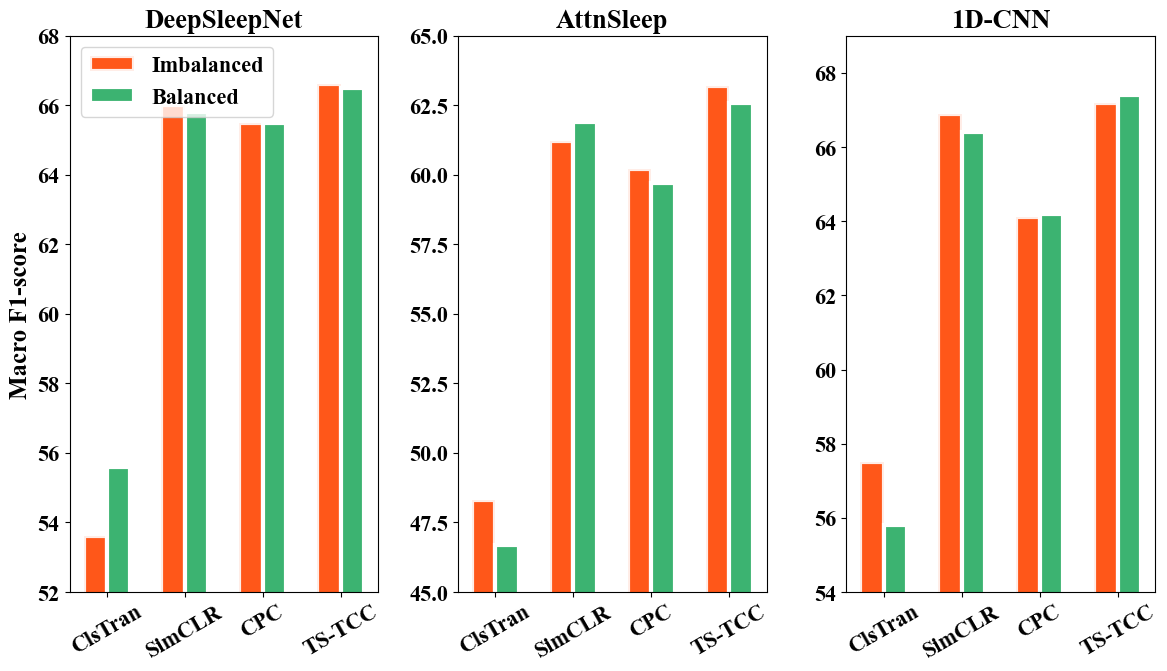}
    \caption{Comparison between pretraining self-supervised algorithms with original imbalanced dataset vs. oversampled original dataset applied on Sleep-EDF dataset.}
    \label{fig:imb_vs_bal}
\end{figure}

The nature of sleep stages implies that some stages, e.g., N1 occur less frequently than other stages such as N2. Consequently, the sleep stage datasets are usually imbalanced (see Table~\ref{tbl:data}). Therefore, it is important to study whether the data imbalance affects the quality of the learned representations by SSL algorithms.
To do so, we compute the performance gap between models pre-trained on balanced and imbalanced datasets. Specifically, we pretrain the SSL algorithms with the original imbalanced data, and also with oversampled balanced data \cite{deepsleepnet}. The experimental results are shown in Fig.~\ref{fig:imb_vs_bal}. 

We observe that the gap between balanced and imbalanced pretraining is minor for \textit{contrastive} SSL algorithms, and it does not exceed a maximum of 0.2\%, 0.6\%, and 0.5\% for DeepSleepNet, AttnSleep, and 1D-CNN respectively. These observations show that contrastive SSL algorithms are more robust to dataset imbalance, which is consistent with previous studies \cite{simclr,iclr_imbalance}. The main reason is their ability to learn more general and richer features from the majority classes than traditional supervised learning. In specific, the learned self-supervised representations are not supervised or motivated by any labels, i.e., they are not label-directed, and they could learn other intrinsic properties in the EEG signal. These features can improve the classification performance of the minor classes and the learned features can be more efficient for the downstream task. On the other hand, the ClsTran algorithm is directed by a cross-entropy loss that depends on the assigned pseudo labels. Therefore, it can be affected by the data imbalance, and it shows different performance with oversampled data. In the supplementary material, we also analyze the ability of SSL to improve the performance of the minor classes (see Section~\ref{sec:supp:minority}).

\subsection{Robustness to Domain-shift}
In some scenarios, we may afford to label the samples of one subject, and we aim to transfer the knowledge from this subject to another unlabeled and out-of-distribution subjects. This distribution shift can be caused by a different data collection methodology or differences in subjects' health status.
To deal with this challenging scenario, some recent works proposed transfer learning and unsupervised domain adaptation algorithms to mitigate the domain shift \cite{adast,phan_tl,slarda,adatime}.

In this section, we investigate the transferability of supervised training against self-supervised pretraining under the domain-shift settings on five random cross-domain (cross-subject) scenarios from Sleep-EDF dataset. Each cross-domain scenario consists of one source subject, and one target subject. The supervised transferability is obtained by training the model on the source domain and directly testing it on the target domain. For the SSL algorithms, we pretrain the model with the unlabeled source data and fine-tune it with its labels, then test it on the target domain. The experimental results are provided in Table~\ref{tbl:transfer_learning}.

We notice that in all the cross-domain scenarios, at least one SSL algorithm can outperform the supervised transferability. However, we notice that the overall (average) improvement is marginal in DeepSleepNet and AttnSleep between supervised and best SSL algorithm with 1.2\% and 1.6\%, respectively. Counterpart, 1D-CNN witnessed an overall improvement of 4.2\%. Considering that 1D-CNN model is less complex than DeepSleepNet and AttnSleep models (see Section~\ref{sec:supp:num_params} in the supplementary materials), we can conclude that SSL can compensate for its lower transferability capacity and allow it to achieve comparable performance.


\section{Discussion \& Recommendations}

In this paper, we studied whether self-supervised pretraining can help improve the performance of existing sleep stage classification  models in the few-labeled data regime. Our experiments were held with four SSL algorithms and applied to three SSC models on three different datasets. The experimental results suggest the following conclusions.
\begin{itemize}
    \item Contrastive SSL algorithms guarantee superior performance of SSC models over supervised training in the few-labeled data settings.

    \item Contrastive SSL algorithms are robust against sleep data imbalance, and this imbalance does not affect the quality of learned representations.
    
    \item Self-supervised pretraining improves the out-of-domain transferability performance in SSC models.
    
    \item SSL with predictive tasks can improve the temporal learning capability of SSC models.
    
\end{itemize}

The above conclusions reveal some potential future works to enhance the SSL algorithms proposed for sleep stage classification. First, we find that the auxiliary task, ClsTran, yields lower performance even than the supervised training in most cases. Therefore, it is important to study the SSC problem and propose a new SSC-specific auxiliary task to be more beneficial to the downstream performance, similar to the proposed tasks in \cite{pmlr-v158-wagh21a}.

Second, our experiments included two contrastive SSL algorithms that rely on data augmentations to chose the positive and negative pairs, i.e., SimCLR and TS-TCC. 
These two methods consider \textit{only} the augmented view of each same sample as the positive pair, and all the other samples are considered negative pairs. 
However, some of these negative pairs may share the same label and semantic information with that anchor sample, and pulling them away from each other may deteriorate the performance. Therefore, one way to improve these algorithms is to reduce the number of false negative samples when applying contrastive learning.
In addition, designing well-suited augmentations for sleep EEG data can learn more effective representations.

Third, SSL algorithms showed limited improvement to the minority classes in the sleep data, i.e., N1 and N3, which limits the overall improvement. 
Therefore, another research direction is to study how can self-supervised algorithms learn more about the characteristics of minority classes during pretraining. Last, we noticed that SSL algorithms had a limited transferability improvement, which can be further investigated.

\section{CONCLUSIONS}
In this paper, we assess the efficacy of different self-supervised learning (SSL) algorithms to improve the performance of sleep stage classification (SSC) models under the few-labels settings. The experimental results reveal that contrastive SSL algorithms can learn more robust and invariant representations about the sleep EEG data. In addition, SSL algorithms that include predictive tasks can learn temporal features about EEG data during pretraining, and hence lessens the need to a temporal encoder in the SSC models. Moreover, self-supervised pretraining can improve the robustness of SSC models against data imbalance and domain shift problems. Hence, we recommend pretraining existing SSC models with contrastive SSL algorithms to become more practical in the real-world label-scarce scenarios.

\bibliographystyle{unsrt}
\bibliography{citations}

\begin{thebibliography}{10}

\bibitem{nature_paper}
I.~Perez-Pozuelo, B.~Zhai, J.~Palotti, R.~Mall, M.~Aupetit, J.~M. Garcia-Gomez,
  S.~Taheri, Y.~Guan, and L.~Fernandez-Luque.
\newblock The future of sleep health: a data-driven revolution in sleep science
  and medicine.
\newblock {\em NPJ digital medicine}, 3(42), 2020.

\bibitem{aasm}
R.~B. Berry, R.~Brooks, C.~Gamaldo, S.~M. Harding, R.~M. Lloyd, S.~F. Quan,
  M.~T. Troester, and B.~V. Vaughn.
\newblock Aasm scoring manual updates for 2017 (version 2.4).
\newblock {\em Journal of clinical sleep medicine}, 13:665–666, 2017.

\bibitem{ssc_timeFreq}
Pankaj Jadhav and Siddhartha Mukhopadhyay.
\newblock Automated sleep stage scoring using time-frequency spectra
  convolution neural network.
\newblock {\em IEEE Transactions on Instrumentation and Measurement}, 71:1--9,
  2022.

\bibitem{ssc_transition}
Jaeun Phyo, Wonjun Ko, Eunjin Jeon, and Heung-II Suk.
\newblock Enhancing contextual encoding with stage-confusion and
  stage-transition estimation for eeg-based sleep staging.
\newblock In {\em IEEE International Conference on Acoustics, Speech and Signal
  Processing (ICASSP)}, pages 1301--1305, 2022.

\bibitem{sleep_transformer}
Huy Phan, Kaare Mikkelsen, Oliver~Y. Chen, Philipp Koch, Alfred Mertins, and
  Maarten De~Vos.
\newblock Sleeptransformer: Automatic sleep staging with interpretability and
  uncertainty quantification.
\newblock {\em IEEE Transactions on Biomedical Engineering}, 69(8):2456--2467,
  2022.

\bibitem{deepsleepnet}
A.~Supratak, H.~Dong, C.~Wu, and Y.~Guo.
\newblock Deepsleepnet: a model for automatic sleep stage scoring based on raw
  single-channel eeg.
\newblock {\em IEEE Transactions on Neural Systems and Rehabilitation
  Engineering}, 25(11):1998--2008, 2017.

\bibitem{attnSleep_paper}
Emadeldeen Eldele, Zhenghua Chen, Chengyu Liu, Min Wu, Chee-Keong Kwoh, Xiaoli
  Li, and Cuntai Guan.
\newblock An attention-based deep learning approach for sleep stage
  classification with single-channel eeg.
\newblock {\em IEEE Transactions on Neural Systems and Rehabilitation
  Engineering}, 29:809--818, 2021.

\bibitem{tstcc}
Emadeldeen Eldele, Mohamed Ragab, Zhenghua Chen, Min Wu, Chee~Keong Kwoh,
  Xiaoli Li, and Cuntai Guan.
\newblock Time-series representation learning via temporal and contextual
  contrasting.
\newblock In {\em Proceedings of the Thirtieth International Joint Conference
  on Artificial Intelligence, {IJCAI-21}}, pages 2352--2359, 2021.

\bibitem{simclr}
Ting {Chen}, Simon {Kornblith}, Mohammad {Norouzi}, and Geoffrey {Hinton}.
\newblock A simple framework for contrastive learning of visual
  representations.
\newblock In {\em International Conference on Machine Learning, {ICML}},
  volume~1, pages 1597--1607, 2020.

\bibitem{cpc}
Aaron van~den {Oord}, Yazhe {Li}, and Oriol {Vinyals}.
\newblock Representation learning with contrastive predictive coding.
\newblock {\em arXiv: Learning}, 2018.

\bibitem{moco}
Kaiming {He}, Haoqi {Fan}, Yuxin {Wu}, Saining {Xie}, and Ross {Girshick}.
\newblock Momentum contrast for unsupervised visual representation learning.
\newblock In {\em 2020 IEEE/CVF Conference on Computer Vision and Pattern
  Recognition (CVPR)}, pages 9729--9738, 2020.

\bibitem{simsiam}
Xinlei {Chen} and Kaiming {He}.
\newblock Exploring simple siamese representation learning.
\newblock In {\em Proceedings of the IEEE/CVF Conference on Computer Vision and
  Pattern Recognition}, pages 15750--15758, 2020.

\bibitem{byol}
Jean-Bastien {Grill}, Florian {Strub}, Florent {Altché}, Corentin {Tallec},
  Pierre~H. {Richemond}, Elena {Buchatskaya}, Carl {Doersch}, Bernardo~Avila
  {Pires}, Zhaohan~Daniel {Guo}, Mohammad~Gheshlaghi {Azar}, Bilal {Piot},
  Koray {Kavukcuoglu}, Rémi {Munos}, and Michal {Valko}.
\newblock Bootstrap your own latent: A new approach to self-supervised
  learning.
\newblock In {\em Advances in Neural Information Processing Systems},
  volume~33, pages 21271--21284, 2020.

\bibitem{coSleep}
Jianan Ye, Qinfeng Xiao, Jing Wang, Hongjun Zhang, Jiaoxue Deng, and Youfang
  Lin.
\newblock Cosleep: A multi-view representation learning framework for
  self-supervised learning of sleep stage classification.
\newblock {\em IEEE Signal Processing Letters}, 29:189--193, 2022.

\bibitem{resnet_sleep}
Ahmed~Imtiaz Humayun, Asif~Shahriyar Sushmit, Taufiq Hasan, and Mohammed
  Imamul~Hassan Bhuiyan.
\newblock End-to-end sleep staging with raw single channel eeg using deep
  residual convnets.
\newblock In {\em 2019 IEEE EMBS International Conference on Biomedical \&
  Health Informatics (BHI)}, pages 1--5, 2019.

\bibitem{iinet}
Hogeon Seo, Seunghyeok Back, Seongju Lee, Deokhwan Park, Tae Kim, and Kyoobin
  Lee.
\newblock Intra- and inter-epoch temporal context network (iitnet) using
  sub-epoch features for automatic sleep scoring on raw single-channel eeg.
\newblock {\em Biomedical Signal Processing and Control}, 61:102037, 2020.

\bibitem{graph}
Qing Cai, Zhongke Gao, Jianpeng An, Shuang Gao, and Celso Grebogi.
\newblock A graph-temporal fused dual-input convolutional neural network for
  detecting sleep stages from eeg signals.
\newblock {\em IEEE Transactions on Circuits and Systems II: Express Briefs},
  68(2):777--781, 2021.

\bibitem{seqsleepnet}
Huy {Phan}, Fernando {Andreotti}, Navin {Cooray}, Oliver~Y. {Chen}, and
  Maarten~De {Vos}.
\newblock Seqsleepnet: End-to-end hierarchical recurrent neural network for
  sequence-to-sequence automatic sleep staging.
\newblock {\em international conference of the ieee engineering in medicine and
  biology society}, 27(3):400--410, 2019.

\bibitem{attn_residual}
Wei Qu, Zhiyong Wang, Hong Hong, Zheru Chi, David~Dagan Feng, Ron Grunstein,
  and Christopher Gordon.
\newblock A residual based attention model for eeg based sleep staging.
\newblock {\em IEEE Journal of Biomedical and Health Informatics},
  24(10):2833--2843, 2020.

\bibitem{puzzle_pretext}
Mehdi Noroozi and Paolo Favaro.
\newblock Unsupervised learning of visual representations by solving jigsaw
  puzzles.
\newblock In {\em ECCV}, 2016.

\bibitem{rotation_pretext}
Spyros Gidaris, Praveer Singh, and Nikos Komodakis.
\newblock Unsupervised representation learning by predicting image rotations.
\newblock In {\em ICML}, 2018.

\bibitem{simclr_like_eeg}
Mostafa~Neo Mohsenvand, Mohammad~Rasool Izadi, and Pattie Maes.
\newblock Contrastive representation learning for electroencephalogram
  classification.
\newblock In {\em Machine Learning for Health NeurIPS Workshop}, 2020.

\bibitem{contrastive_ssc}
Xue Jiang, Jianhui Zhao, Bo~Du, and Zhiyong Yuan.
\newblock Self-supervised contrastive learning for eeg-based sleep staging.
\newblock In {\em International Joint Conference on Neural Networks (IJCNN)},
  pages 1--8, 2021.

\bibitem{uncovering_eeg}
Hubert Banville, Omar Chehab, Aapo Hyvarinen, Denis Engemann, and Alexandre
  Gramfort.
\newblock Uncovering the structure of clinical eeg signals with self-supervised
  learning.
\newblock {\em Journal of Neural Engineering}, 2020.

\bibitem{ssl_pred_disc}
Qinfeng Xiao, Jing Wang, Jianan Ye, Hongjun Zhang, Yuyan Bu, Yiqiong Zhang, and
  Hao Wu.
\newblock Self-supervised learning for sleep stage classification with
  predictive and discriminative contrastive coding.
\newblock In {\em IEEE International Conference on Acoustics, Speech and Signal
  Processing (ICASSP)}, pages 1290--1294, 2021.

\bibitem{ijcai2022p537}
Harim Lee, Eunseon Seong, and Dong-Kyu Chae.
\newblock Self-supervised learning with attention-based latent signal
  augmentation for sleep staging with limited labeled data.
\newblock In Lud~De Raedt, editor, {\em Proceedings of the Thirty-First
  International Joint Conference on Artificial Intelligence, {IJCAI-22}}, pages
  3868--3876. International Joint Conferences on Artificial Intelligence
  Organization, 7 2022.
\newblock Main Track.

\bibitem{SleepECL}
Hongjun Zhang, Jing Wang, Jiahong Xiong, Yuxuan Ding, Zhenliang Gan, and
  Youfang Lin.
\newblock Expert knowledge inspired contrastive learning for sleep staging.
\newblock In {\em 2022 International Joint Conference on Neural Networks
  (IJCNN)}, pages 1--6, 2022.

\bibitem{har_ssl}
Aaqib Saeed, Tanir Ozcelebi, and Johan Lukkien.
\newblock Multi-task self-supervised learning for human activity detection.
\newblock {\em Proc. ACM Interact. Mob. Wearable Ubiquitous Technol.}, 3(2),
  2019.

\bibitem{ecg_emotion_rec}
P.~Sarkar and A.~Etemad.
\newblock Self-supervised ecg representation learning for emotion recognition.
\newblock {\em IEEE Transactions on Affective Computing}, 2020.

\bibitem{emadeldeen2022catcc}
Emadeldeen Eldele, Mohamed Ragab, Zhenghua Chen, Min Wu, Chee~Keong Kwoh,
  Xiaoli Li, and Cuntai Guan.
\newblock Self-supervised contrastive representation learning for
  semi-supervised time-series classification.
\newblock {\em arXiv preprint arXiv:2208.06616}, 2022.

\bibitem{ssl_survey}
Ashish Jaiswal, Ashwin~Ramesh Babu, Mohammad~Zaki Zadeh, Debapriya Banerjee,
  and Fillia Makedon.
\newblock A survey on contrastive self-supervised learning.
\newblock {\em Technologies}, 9(1), 2021.

\bibitem{sleep_edf}
Ary~L Goldberger, Luis~AN Amaral, Leon Glass, Jeffrey~M Hausdorff, Plamen~Ch
  Ivanov, Roger~G Mark, Joseph~E Mietus, George~B Moody, Chung-Kang Peng, and
  H~Eugene Stanley.
\newblock Physiobank, physiotoolkit, and physionet: components of a new
  research resource for complex physiologic signals.
\newblock {\em circulation}, 101(23):e215--e220, 2000.

\bibitem{shhs_ref1}
G.~Zhang, L.~Cui, R.~Mueller, S.~Tao, M.~Kim, M.~Rueschman, S.~Mariani,
  D.~Mobley, and S.~Redline.
\newblock The national sleep research resource: towards a sleep data commons.
\newblock {\em Journal of the American Medical Informatics Association},
  25(10):1351--1358, 2018.

\bibitem{shhs_ref2}
S.~F. Quan, B.~V. Howard, C.~Iber, J.~P. Kiley, F.~J. Nieto, G.~T. O'Connor,
  D.~M. Rapoport, S.~Redline, J.~Robbins, J.~M. Samet, et~al.
\newblock The sleep heart health study: design, rationale, and methods.
\newblock {\em Sleep}, 20(12):1077--1085, 1997.

\bibitem{isruc_dataset}
Sirvan Khalighi, Teresa Sousa, José~Moutinho Santos, and Urbano Nunes.
\newblock Isruc-sleep: A comprehensive public dataset for sleep researchers.
\newblock {\em Computer Methods and Programs in Biomedicine}, 124:180--192,
  2016.

\bibitem{9157100}
Alejandro Newell and Jia Deng.
\newblock How useful is self-supervised pretraining for visual tasks?
\newblock In {\em 2020 IEEE/CVF Conference on Computer Vision and Pattern
  Recognition (CVPR)}, pages 7343--7352, 2020.

\bibitem{iclr_imbalance}
Hong Liu, Jeff~Z. HaoChen, Adrien Gaidon, and Tengyu Ma.
\newblock Self-supervised learning is more robust to dataset imbalance.
\newblock In {\em International Conference on Learning Representations}, 2022.

\bibitem{adast}
Emadeldeen Eldele, Mohamed Ragab, Zhenghua Chen, Min Wu, Chee-Keong Kwoh,
  Xiaoli Li, and Cuntai Guan.
\newblock Adast: Attentive cross-domain eeg-based sleep staging framework with
  iterative self-training.
\newblock {\em IEEE Transactions on Emerging Topics in Computational
  Intelligence}, pages 1--12, 2022.

\bibitem{phan_tl}
Huy Phan, Oliver~Y. Chén, Philipp Koch, Zongqing Lu, Ian McLoughlin, Alfred
  Mertins, and Maarten De~Vos.
\newblock Towards more accurate automatic sleep staging via deep transfer
  learning.
\newblock {\em IEEE Transactions on Biomedical Engineering}, 68(6):1787--1798,
  2021.

\bibitem{slarda}
Mohamed Ragab, Emadeldeen Eldele, Zhenghua Chen, Min Wu, Chee-Keong Kwoh, and
  Xiaoli Li.
\newblock Self-supervised autoregressive domain adaptation for time series
  data.
\newblock {\em IEEE Transactions on Neural Networks and Learning Systems},
  pages 1--11, 2022.

\bibitem{adatime}
Mohamed Ragab, Emadeldeen Eldele, Wee~Ling Tan, Chuan-Sheng Foo, Zhenghua Chen,
  Min Wu, Chee-Keong Kwoh, and Xiaoli Li.
\newblock Adatime: A benchmarking suite for domain adaptation on time series
  data.
\newblock {\em arXiv preprint arXiv:2203.08321}, 2022.

\bibitem{pmlr-v158-wagh21a}
Neeraj Wagh, Jionghao Wei, Samarth Rawal, Brent Berry, Leland Barnard, Benjamin
  Brinkmann, Gregory Worrell, David Jones, and Yogatheesan Varatharajah.
\newblock Domain-guided self-supervision of eeg data improves downstream
  classification performance and generalizability.
\newblock In {\em Proceedings of Machine Learning for Health}, volume 158 of
  {\em Proceedings of Machine Learning Research}, pages 130--142. PMLR, 04 Dec
  2021.

\end{thebibliography}

\clearpage
\newpage
\title{Supplementary Materials of: \\ Self-supervised Learning for label-efficient Sleep Stage Classification: A Comprehensive Evaluation}

\maketitle

\renewcommand\thefigure{S.\arabic{figure}}
\renewcommand{\thesection}{S.\Roman{section}} 
\renewcommand\thetable{S.\arabic{table}}
\setcounter{figure}{0}
\setcounter{table}{0}
\setcounter{section}{0}

\section{Number of Parameters}
\label{sec:supp:num_params}
Table~\ref{tbl:num_params} shows the number of parameters of the three adopted sleep stage classification models.
We notice that DeepSleepNet has the highest number of trainable parameters, while 1D-CNN has the lowest.
Notably, we only use the feature extractor in the self-supervised pretraining, and its complexity is an important factor with respect to the SSL algorithm.

\begin{table}[!htb]
\centering
\caption{Comparison between the number of trainable parameters in the three adopted SSC models.}
\begin{tabular}{@{}lcccc@{}}
\toprule
             & Feat. Ext. & Temporal Enc. & Classifier & Total      \\ \midrule
DeepSleepNet & 604,928           & 35,275,776       & 5,125      & 35,885,829 \\
AttnSleep    & 407,600           & 103,200          & 12,005     & 522,805    \\
1D-CNN        & 83,168            & 0                & 41,605     & 124,773    \\ \bottomrule
\end{tabular}
\label{tbl:num_params}
\end{table}

\section{Data Augmentations}
\label{sec:data_augmentation}
In this work, some SSL algorithms need data augmentations to perform. Therefore, we examine four augmentations, as follows.

\begin{itemize}
    \item  Noise: includes adding a randomly generated noise signal with a mean of 0 and standard deviation of 0.8.
    
    \item Time shift: shifting the signal with 20\% of the total signal timesteps and rotating the shifted part back to the beginning of the signal.
    
    
    \item Negate: multiply the value of the signal by a factor of -1.
    
    \item Permute: randomly split each signal into five segments in the time domain, then permute the segments, and re-combine them into their original shape.
\end{itemize}

The ClsTran algorithm applies these four augmentations to classify between them. For SimCLR and TS-TCC, we used their corresponding augmentations as applied in \cite{tstcc}.

\section{Experiments}

\subsection{Does SSL improve the performance of minority classes?}
\label{sec:supp:minority}
\begin{figure}
    \centering
    \subfigure[DeepSleepNet]{\label{fig:ssl_comp_dsn_edf}\includegraphics[width=0.96\columnwidth]{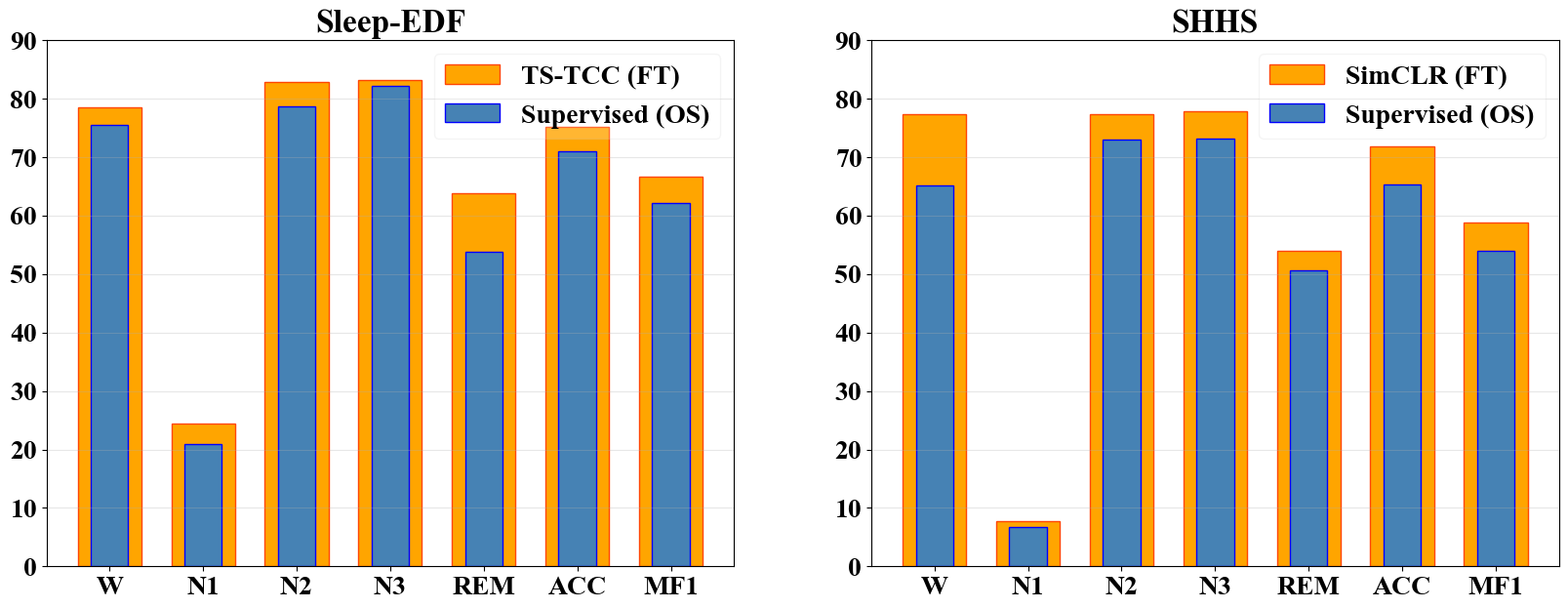}}
    \subfigure[AttnSleep]{\label{fig:ssl_comp_attn_edf}\includegraphics[width=0.96\columnwidth]{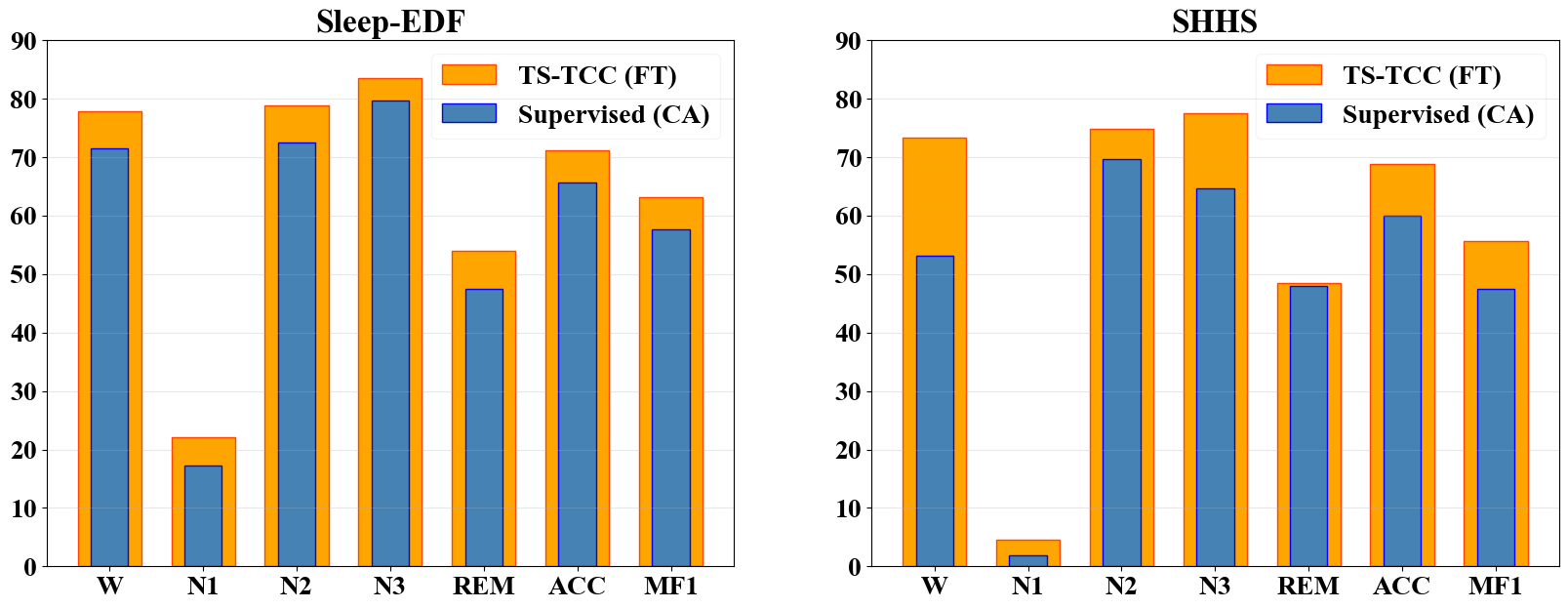}}

    \caption{Per-class performance (in terms of F1-score) comparison between SSC models when trained with their specific methods for addressing class-imbalance problem and best-performing SSL algorithm. (a) DeepSleepNet with oversampling (OS) pretraining, (b) AttnSleep with class-aware (CA) loss function.}
    \label{fig:imb_ssl_comp}
\end{figure}

The second question is about the ability of SSL algorithms to improve the performance of the minority classes, e.g., N1 stage. This is an emerging problem in SSC works, which usually aim to find a way to improve the N1 class performance.
Therefore, we compare the performance of self-supervised pretraining with the SSC models including their proposed techniques to address the class-imbalance problem.
Fig.~\ref{fig:imb_ssl_comp} shows the comparison results, where DeepSleepNet is trained in a two-stage procedure, i.e., oversampled-data pretraining then fine-tuning, and AttnSleep is trained with including its class-aware loss function.

The results show that SSL algorithms surpass the techniques proposed by both works, as they can achieve better performance across all classes. 
However, our focus here is on the minor classes, i.e., N1 and N3. We find that SSL algorithms do not improve their performance significantly, which opens a new research direction to find a way to improve the SSL performance in these classes.

\subsection{Performance Under Different Few-labels Settings}
\label{sec:supp:few_lbls}
\begin{figure}[!tb]
    \centering
    \includegraphics[width=\columnwidth]{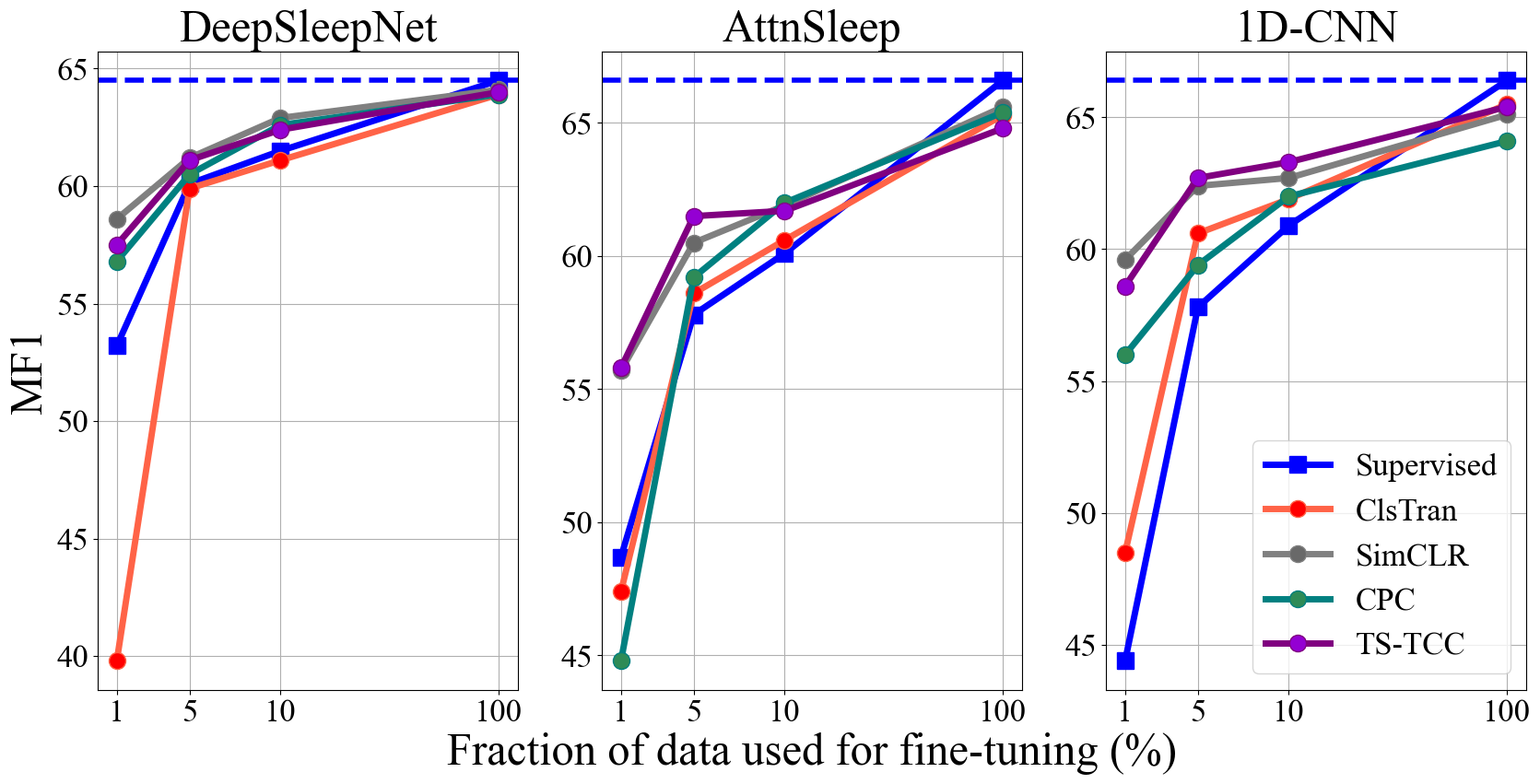}
    \caption{Fine-tuning the pretrained SSL algorithms with different fractions of labeled SHHS data.}
    \label{fig:few_lbl_shhs}
\end{figure}

\begin{figure}[!tb]
    \centering
    \includegraphics[width=\columnwidth]{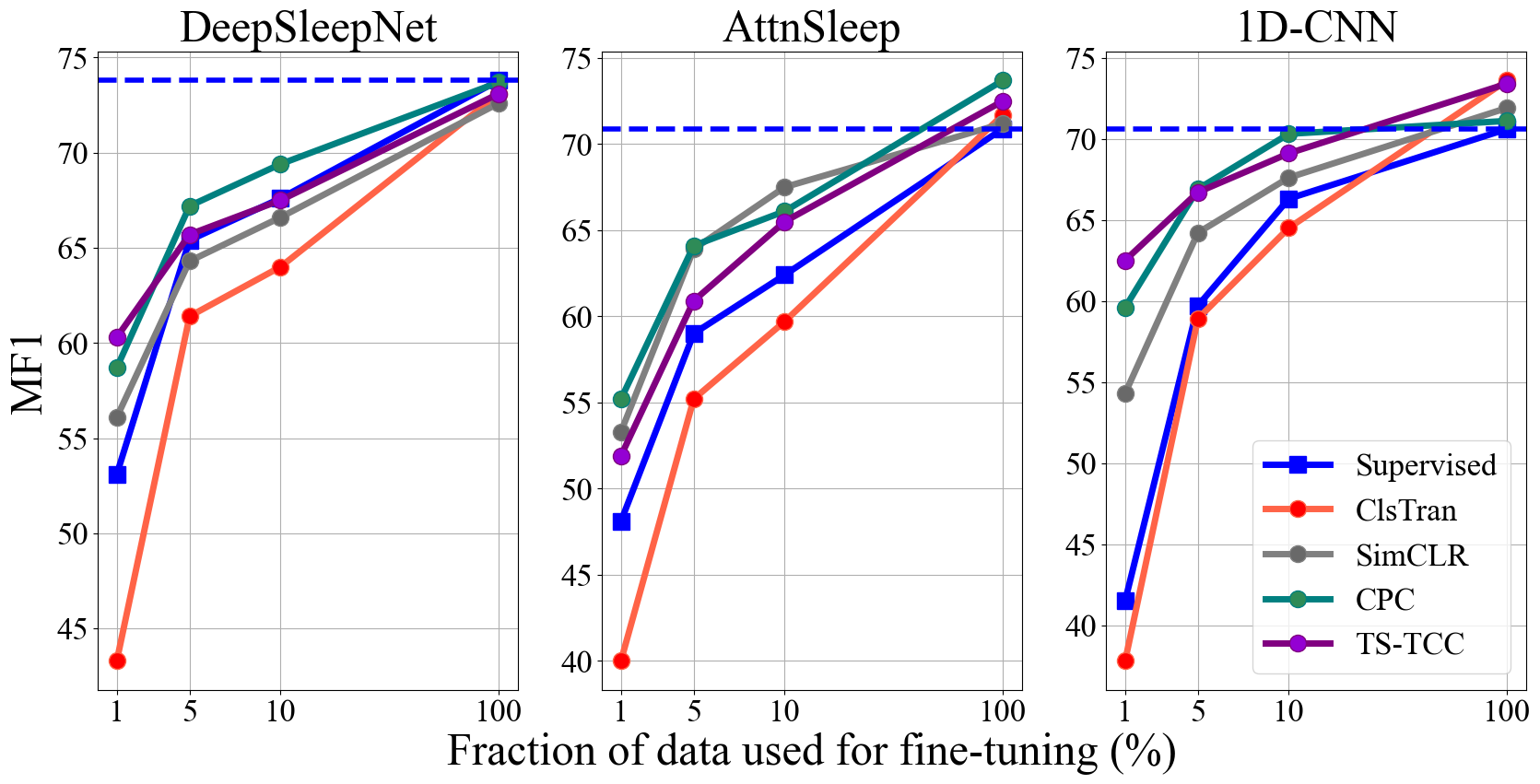}
    \caption{Fine-tuning the pretrained SSL algorithms with different fractions of labeled ISRUC data.}
    \label{fig:few_lbl_isruc}
\end{figure}

We discuss the results of pretrained models when fine-tuned with different amounts of labeled data for both SHHS and ISRUC datasets, as shown in Fig.~\ref{fig:few_lbl_shhs}~and~\ref{fig:few_lbl_isruc}.
We notice that similar to the conclusions drawn from Sleep-EDF dataset, we find that with 5 or 10\% of labels, we can achieve very close performance to the supervised training with full labels.
In addition, by fine-tuning with the full labels, we can surpass the fully supervised training on the ISRUC dataset in AttnSleep and 1D-CNN.


\begin{figure}
    \centering
    \subfigure[DeepSleepNet]{\label{fig:te_dsn}\includegraphics[width=0.96\columnwidth]{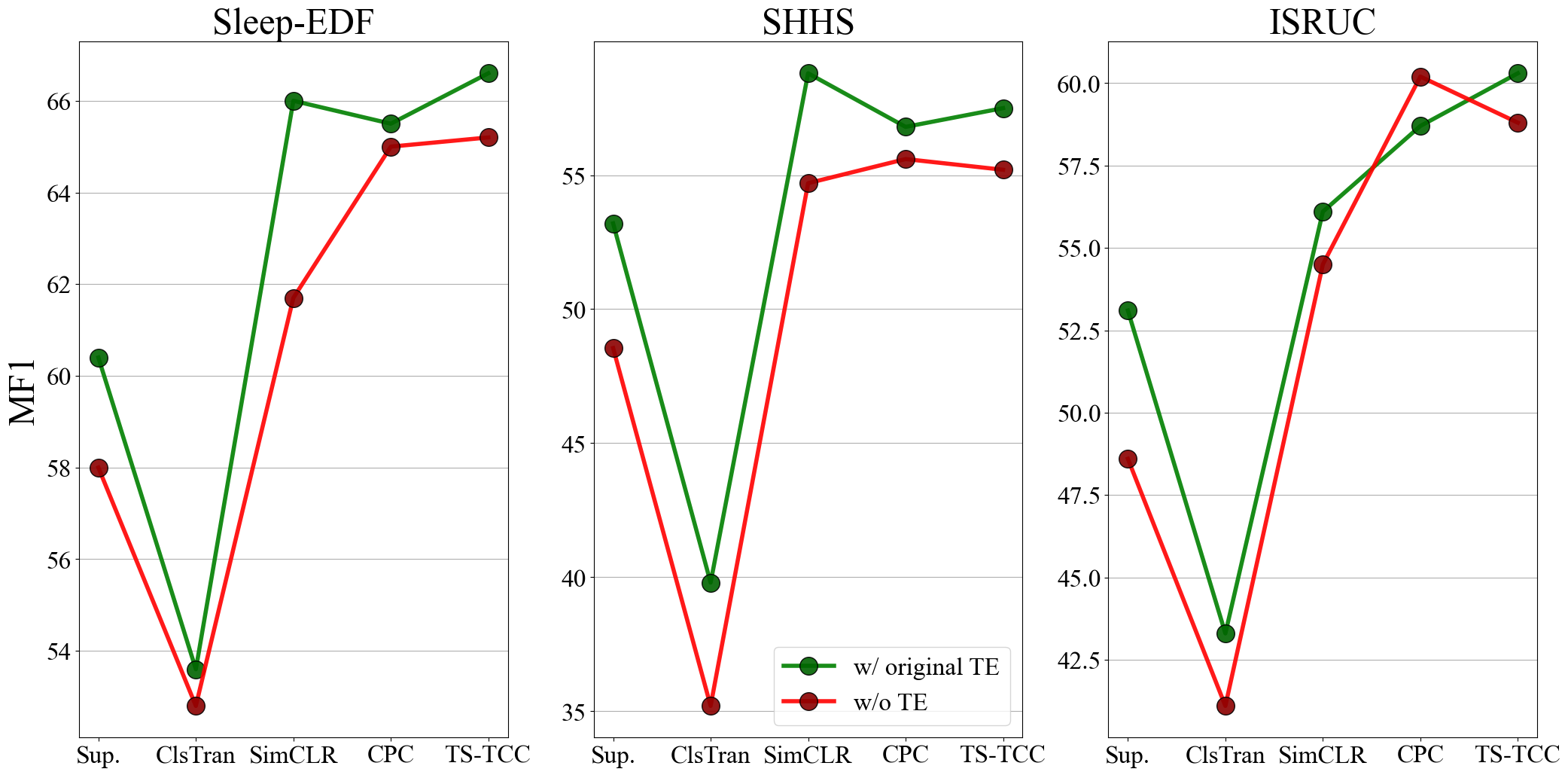}}
    \subfigure[AttnSleep]{\label{fig:te_attn}\includegraphics[width=0.96\columnwidth]{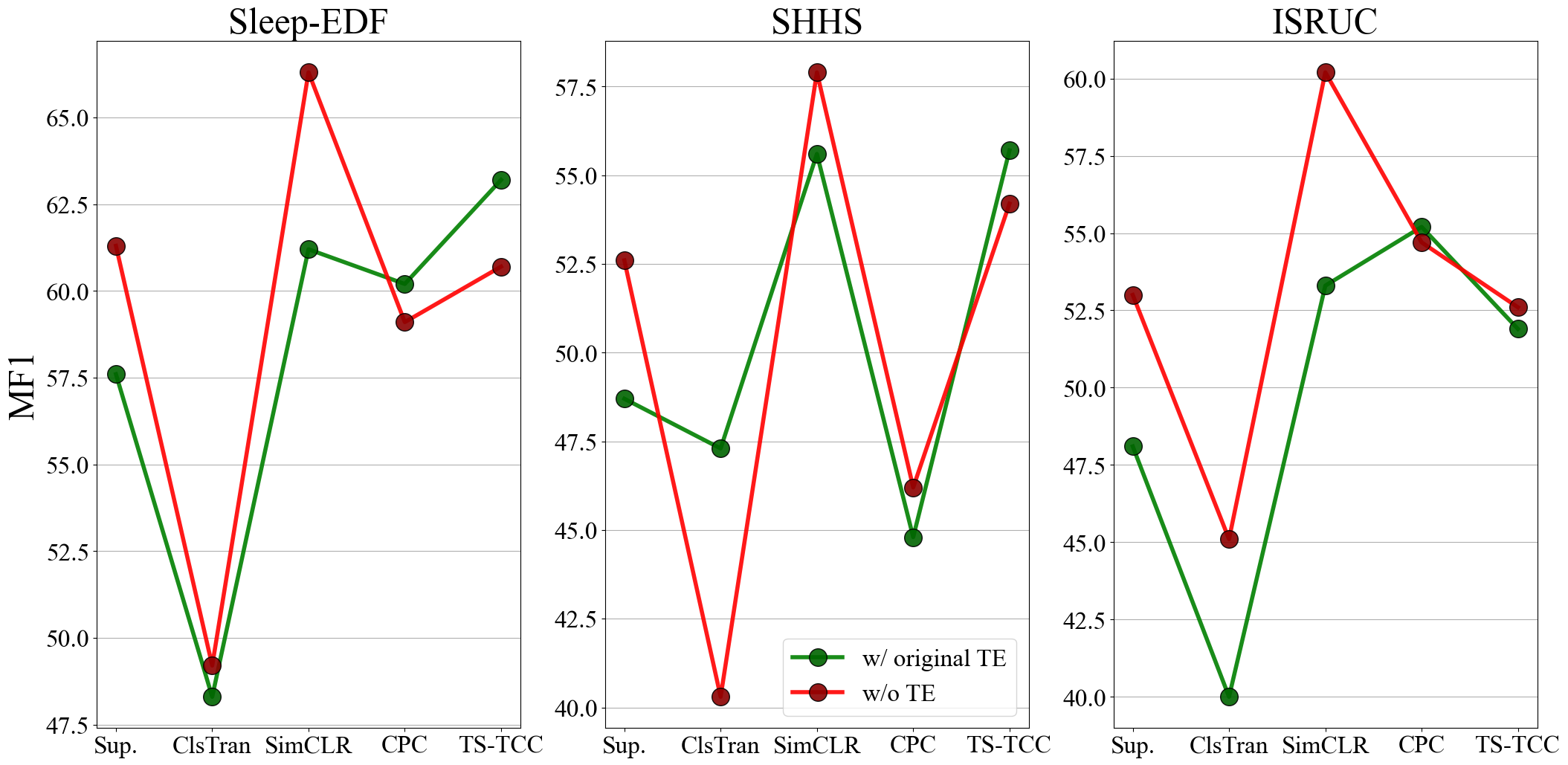}}

    \caption{Performance of SSC models when fine-tuned with and without temporal encoder. The performance of SSL models that pretrain by predicting the future timesteps, i.e., CPC and TS-TCC, is more robust to the existence of temporal encoders.}
    \label{fig:fine_tune_te}
\end{figure}
\begin{figure}
    \centering
    \subfigure[DeepSleepNet]{\label{fig:te_dsn_swapped}\includegraphics[width=0.96\columnwidth]{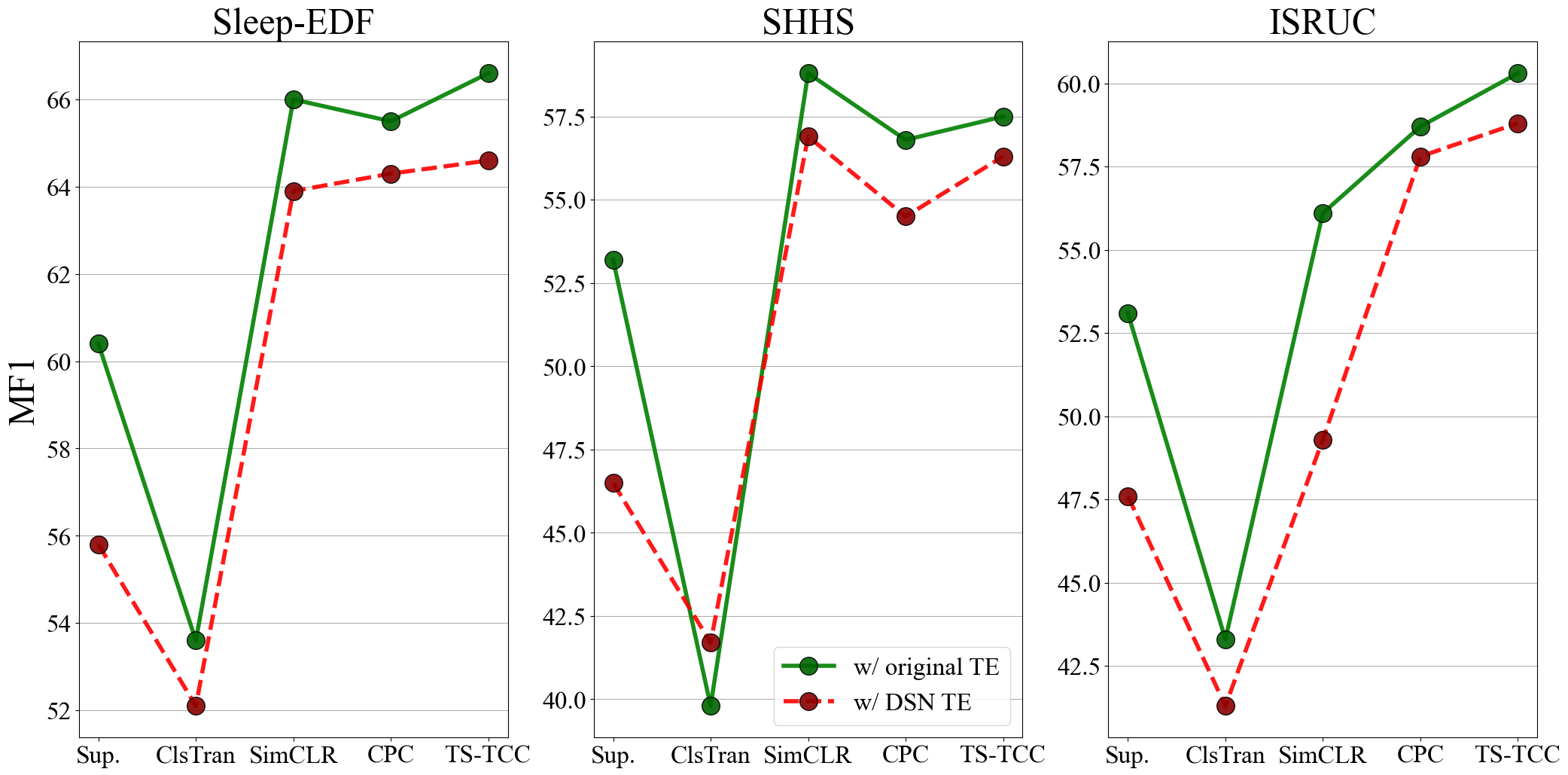}}
    \subfigure[AttnSleep]{\label{fig:te_attn_swapped}\includegraphics[width=0.96\columnwidth]{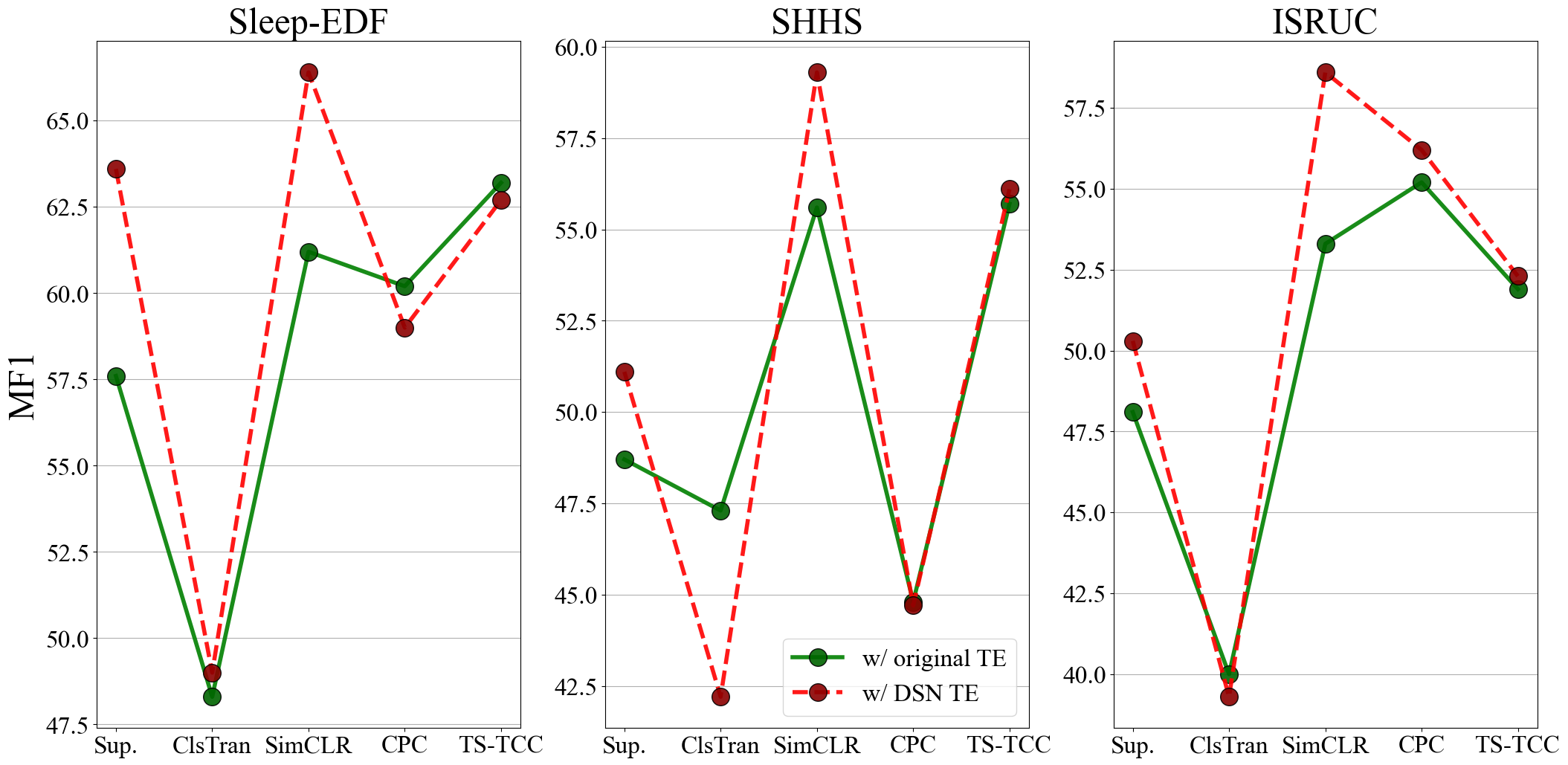}}

    \caption{Performance of SSC models when fine-tuned with original Temporal Encoder and with different Temporal Encoder proposed for the other method. Here, AttnSleep is fine-tuned with BiLSTM used in DeepSleepNet. Also, DeepSleepNet is fine-tuned with causal self-attention used in AttnSleep.}
    \label{fig:fine_tune_swaped_te}
\end{figure}

\subsection{Can SSL compensate temporal encoder?}
\label{sec:supp:te}
One of the main questions when designing a sleep stage classification model is how to learn the temporal dependencies in EEG data. Therefore, we study the capability of SSL to learn and characterize the temporal dependencies.
Specifically, we examine the performance of pretrained SSC models by fine-tuning them with and without the temporal encoders. The experimental results are visualized in Fig.~\ref{fig:fine_tune_te}.
In addition, we examine their robustness to the type of the temporal encoder by fine-tuning DeepSleepNet with the causal self-attention used in AttnSleep, and fine-tuning AttnSleep with BiLSTM used in DeepSleepNet, as shown in Fig.~\ref{fig:fine_tune_swaped_te}.

We find that supervised training is more affected by the existence and the type of the temporal encoder, as it shows unstable performance with changing these factors. 
Second, we find that CPC and TS-TCC are more robust to the existence and the type of the temporal encoder regardless of the SSC model or the dataset used. 
The reason is that these two approaches learn representations by predicting the future timesteps with an autoregressive model. This allows the model to learn the temporal dependencies in the EEG data during the pretraining. Therefore, fine-tuning models pretrained with these two algorithms become more robust against the type or the existence of a temporal encoder.
In contrast, ClsTran and SimCLR mainly rely on data augmentations, which help more to learn spatial representations. Therefore, we find that these two methods are more affected by the type or existence of the temporal encoder.

\end{document}